\documentclass[]{aa} % for a referee version
\usepackage{graphicx, textcomp}
\usepackage{multirow}
\usepackage{natbib}
\usepackage[switch]{lineno}
\usepackage{txfonts}
 \usepackage{hyperref}
 \hypersetup{
      colorlinks   = true,
      citecolor    = blue
 }

\newcommand{\dg}{$^\circ$}
\newcommand{\g}{$\gamma$}

\begin{document} 

\title{The TeV-emitting radio galaxy 3C\,264}
\subtitle{VLBI kinematics and SED modeling}
 \author{B. Boccardi\inst{1,2}
 \and G. Migliori\inst{3}
 \and P. Grandi \inst{2}
 \and E. Torresi\inst{2,4}
 \and F. Mertens \inst{5}
 \and V. Karamanavis \inst{6}
 \and R. Angioni \inst{1,7}
 \and C. Vignali \inst{4,8}
}

\institute{Max-Planck-Institut f\"{u}r Radioastronomie, Auf dem H\"{u}gel 69, D-53121 Bonn, Germany \and INAF -- Osservatorio di Astrofisica e Scienza dello Spazio di Bologna, Via Gobetti 101, I-40129 Bologna, Italy \and INAF – Istituto di Radioastronomia, Bologna, Via Gobetti 101, I-40129 Bologna, Italy  \and  Dipartimento di Fisica e Astronomia, Universit\`a degli Studi di Bologna, Via Gobetti 93/2, I-40129 Bologna, Italy \and Kapteyn Astronomical Institute, University of Groningen, P. O. Box 800, 9700 AV Groningen, The Netherlands \and Fraunhofer-Institut für Hochfrequenzpqhysik und Radartechnik FHR, Fraunhoferstra{\ss}e 20, 53343 Wachtberg, Germany \and Institut f\"ur Theoretische Physik und Astrophysik, Universit\"at W\"urzburg, Emil-Fischer-Str. 31, 97074 W\"urzburg, Germany \and INAF -- Osservatorio di Astrofisica e Scienza dello Spazio di Bologna, Via Gobetti 93/3, I-40129 Bologna, Italy} 

\date{Received 31 January 2019 / Accepted 15 May 2019}

\abstract
{In March 2018, \cite{2018ATel11436....1M} reported the detection by VERITAS of very-high-energy emission ($VHE; > 100$\,${\rm GeV}$) from  3C\,264. This is the sixth, and second most distant, radio galaxy ever detected in the TeV regime.}
{In this article we present a radio and X-ray analysis of the jet in 3C\,264. We determine the main physical parameters of the parsec-scale flow and explore the implications of the inferred kinematic structure for radiative models of this $\gamma$-ray emitting jet.}
{The radio data set is comprised of VLBI observations at 15 GHz from the MOJAVE program, and cover a time period of about two years. Through a segmented wavelet decomposition method (WISE code) we estimate the apparent displacement of individual plasma features; we then perform a pixel-based analysis of the stacked image to determine the jet shape. The X-ray data set includes all available observations from the Chandra, XMM, and Swift satellites, and is used, together with archival data in the other bands, to build the SED.}
{Proper motion is mostly detected along the edges of the flow, which appears strongly limb-brightened. The apparent speeds increase as a function of distance from the core up to a maximum of ${\sim}11.5\,\rm{c}$. This constrains the jet viewing angle to assume relatively small values ($\theta\lesssim10^{\circ}$). In the acceleration region, extending up to a de-projected distance of ${\sim}4.8\times10^4$ Schwarzschild radii (${\sim}11$\,$\rm pc$), the jet is collimating ($r\propto z^{0.40\pm 0.04}$), as predicted for a magnetically-driven plasma flow. By assuming that the core region is indeed magnetically dominated ($U_B/U_e>1$), the SED and the jet power can be well reproduced in the framework of leptonic models, provided that the high-energy component is associated to a second emitting region. The possibility that this region is located at the end of the acceleration zone, either in the jet layer or in the spine, is explored in the modeling.}
{}

   \maketitle

%
%________________________________________________________________
\section{Introduction}
Over the past 10 years, the gamma-ray sky has been thoroughly probed by the {\it Fermi}-LAT space telescope, and active galactic nuclei (AGN) can now be firmly defined as the dominant class among $\gamma$-ray emitters \citep{2015ApJ...810...14A}. In most cases, the $\gamma$-ray detections have been associated to jetted-AGN seen at small angles, i.e., to blazars, while only ${\sim}2$\% of the LAT-detected AGN are classified as radio galaxies. The low detection rate of radio galaxies is due to the relativistic nature of plasma flows in AGN: the high-energy emission is preferentially detected when amplified by relativistic Doppler boosting, i.e., when the narrow boosting cone intercepts our line-of-sight. \\
Radio galaxies which are detected at very high energies ($VHE; > 100$\,${\rm GeV}$) with Cherenkov telescopes are, as expected, even rarer. To date, the class of TeV-emitting AGN comprises only 6 misaligned jets,\footnote{http://tevcat.uchicago.edu/}: Centaurus\,A, M\,87, 3C\,84, IC\,310, PKS\,0625-35, and since March 2018 \citep{2018ATel11436....1M}, 3C\,264, subject of the present study. These nearby galaxies all share radio morphologies and energetics typical of Fanaroff-Riley I sources (FRIs), although IC\,310 and PKS\,0625-35 may be rather classified as intermediate FRI-BL Lac sources \citep{2012A&A...538L...1K, 2018MNRAS.476.4187H}. The newly detected 3C\,264 is at present the second most distant TeV-emitting radio galaxy, being located at a redshift $z=0.0217$. 

Modeling the high-energy emission from radio galaxies, while highly instructive for testing the unified scheme, as well as our understanding of the $\gamma$-ray emission mechanism in AGN, has shown to be challenging. One zone leptonic models, the synchrotron self-Compton (SSC) \citep{1992ApJ...397L...5M, 1996ApJ...461..657B} and the external Compton (EC) \citep{1993ApJ...416..458D, 1994ApJ...421..153S, 2000ApJ...545..107B,2001MNRAS.321L...1C}, are often adequate to reproduce the broadband spectral energy distribution (SED) in blazars \citep[e.g.,][]{2013ApJ...768...54B}, but appear insufficient for TeV BL Lacs and for radio galaxies \citep[e.g.,][]{2005ApJ...634L..33G, 2005ApJ...630..130B, 2011MNRAS.413.2785B, 2014A&A...562A..12P}. One of the main difficulties is that the Lorentz factors $\Gamma$ and Doppler factors $\delta$ inferred from radio studies for these two classes are usually low ($\Gamma,\delta\lesssim 5$) \citep{1999APh....11...19M, 2004ApJ...600..115P, 2010ApJ...723.1150P, 2016AJ....152...12L,2017ApJ...846...98J}, too low to account for the high-energy component of the SEDs. Several alternative models have been proposed that relax the assumption of a one-zone, homogeneous emission region \citep[see, e.g.,][for a recent review]{2018Galax...6..116R}. For instance, it has been shown that if velocity gradients exist in the flow, either along the jet axis  \cite[see, e.g., the decelerating jet model in][]{2003ApJ...594L..27G} or transverse to it \citep[see, e.g., the spine-sheath model in][]{2005A&A...432..401G}, the inverse Compton emission is enhanced by the interaction between the different zones. The existence of transverse velocity gradients is supported by observations, e.g., by VLBI studies of the nearby jets in Mrk 501 \citep{2004ApJ...600..127G}, 3C\,84 \citep{2014ApJ...785...53N, 2019A&A...622A.196K, 2018NatAs...2..472G}, M\,87 \citep{2016A&A...595A..54M, 2018A&A...616A.188K} and Cygnus\,A \citep{2016A&A...585A..33B}, and the spine-sheath scenario has been successfully applied to reproduce the TeV emission in 3C\,84 \citep{2014MNRAS.443.1224T}. Deceleration of FRI jets on sub-kiloparsec and kiloparsec scales is likewise supported by abundant observational evidence \citep[e.g.,][]{1997MNRAS.288L...1H, 1999A&A...341...29F,2002MNRAS.336.1161L}, although its role in the TeV production may be challenged by the fast variability that frequently characterize this emission, implying it originates in very compact regions likely located in the innermost jet.

In addition to the multi-zone leptonic models, hadronic and lepto-hadronic models \citep[e.g.,][]{1992A&A...253L..21M, 1997ApJ...478L...5D,2005ApJ...621..176B, 2013MNRAS.434.2684M}, which attribute the high-energy SED component to proton-synchrotron or to proton-photon interactions, provide valid alternative solutions for reproducing the TeV emission in blazars and in radio galaxies \citep{2000NewA....5..377A, 2003APh....18..593M,2005A&A...442..895A, 2003APh....19..559P,2014A&A...562A..12P}. The hypothesis that relativistic protons also exist in AGN jets has been recently corroborated by the identification of several BL Lacs as possible sources of the neutrinos detected by IceCube \citep{2014MNRAS.443..474P, 2016NatPh..12..807K, 2018Sci...361..147I}, since neutrinos are a natural secondary product of proton-photon interactions. 

In this article we investigate the radio and X-ray properties of the jet in 3C\,264, and we use the inferred parameters to model the broadband SED. This radio source is well studied on sub-kiloparsec and larger scales, but the innermost regions of its jet, where the most energetic processes are likely to take place, are still largely unexplored. Through multi-epoch VLBI observations, we examine the kinematic properties, the internal structure and the geometry of the plasma flow, which are fundamental parameters in the investigation of the high-energy emission mechanisms. The paper is organized as follows. In Section 2 we summarize previous studies of the source. In Section 3 we present the radio and X-ray data set. In Section 4 we describe the methods adopted in the analysis of the VLBI images and we discuss the results. In Section 5 we model the broadband SED. Throughout the article we assume a $\Lambda$CDM cosmology with H$_\mathrm{0}$= 70.5 h$^{-1}$ km s $^{-1}$ Mpc $^{-1}$, $\Omega_\mathrm{M}=0.27$, $\Omega_{\mathrm{\Lambda}}=0.73$ \citep{H0}.

\section{The radio galaxy 3C\,264}
3C\,264 is a nearby ($D_{\rm L}=94$\,$\rm Mpc$) Fanaroff-Riley I radio galaxy whose jet has been extensively studied in the radio, optical, and X-ray bands \citep[][and references therein]{1997ApJ...474..179L, 1997ApJ...483..178B, 2004A&A...415..905L, 2010ApJ...708..171P, 2015Natur.521..495M}. The active source is hosted by the elliptical galaxy NGC\,3862, located off-center in the rich cluster Abell\,1367, and harbouring a supermassive black hole with mass $M_{\rm BH}\sim4.7\times10^8$\,$M_{\odot}$. This mass was derived by \cite{2008A&A...486..119B} from the stellar velocity dispersion using the relation by \cite{2002ApJ...574..740T}. On kilo-parsec scales, 3C\,264 presents a complex head-tailed radio structure embedded in a vast region of low surface brightness emission, which may be indicative of a dense environment. The jet and the weak counter-jet emanate from a compact core, and extend respectively in the north-east and south-west direction. In the inner 2.2 arcseconds, the approaching jet is also shining at optical and X-ray energies via the synchrotron mechanism \citep{2010ApJ...708..171P}. The analysis presented by \cite{2010ApJ...708..171P} also showed that the radio and the optical jet are highly polarized on sub-kiloparsec scales, with some regions reaching fractional polarization of the order of 50\%. The magnetic field vectors are remarkably well aligned with the jet axis, both at the jet boundary and at the center. 

The jet orientation with respect to the observer is uncertain. {\it Hubble Space Telescope} imaging reveals a jet punctuated by few bright knots, apparently emanating from a circular optical ring of unclear nature (see the discussion below). The fastest optical knot was observed to propagate with an apparent speed $\beta_{\rm app}$\,${\sim}7$\,$\rm c$, and to ¨catch up¨ with a slower moving knot  ($\beta_{\rm app}$\,${\sim}1.8$\,$\rm c$) located farther downstream, and possible site of particles acceleration \citep{2015Natur.521..495M}. The kinematic measurement of the maximum apparent speed constrains the jet viewing angle $\theta$ to assume relatively small values ($\theta\lesssim16^{\circ}$). Based on the analysis of the longitudinal brightness profiles and of the jet-to-counter jet ratio, previous radio studies had determined a rather different jet orientation of ${\sim}50^{\circ}$ \citep{1997ApJ...483..178B, 2004A&A...415..905L}, which is however difficult to reconcile with the remarkably circular appearance of the optical ring, seen at a projected radius of ${\sim}300-400$\,$\rm pc$. In the simplest scenario, the ring marks the boundary of a flattened dust disk seen face-on, as observed in several FRI galaxies with optical jets \citep{2000ApJ...542..667S}; therefore the jet axis, which is most likely perpendicular to the disk, should not be oriented at a very large angle. 
More complex interpretations of the ring (a puffed-up disk or a spherical bubble) have been proposed by \cite{1997ApJ...483..178B} and \cite{2004A&A...415..905L} in view of the fact that its location approximately coincides with the region where the jet starts deflecting towards the north, expands faster and becomes dimmer. These elements would suggest an actual physical interaction between the ring and the jet, and thus an alternative geometry.

\section{Data set}
\subsection{Radio data}

\begin{figure}
\centering 
\includegraphics[trim=0cm 0cm 0cm 0cm, clip=true, width=0.47\textwidth]{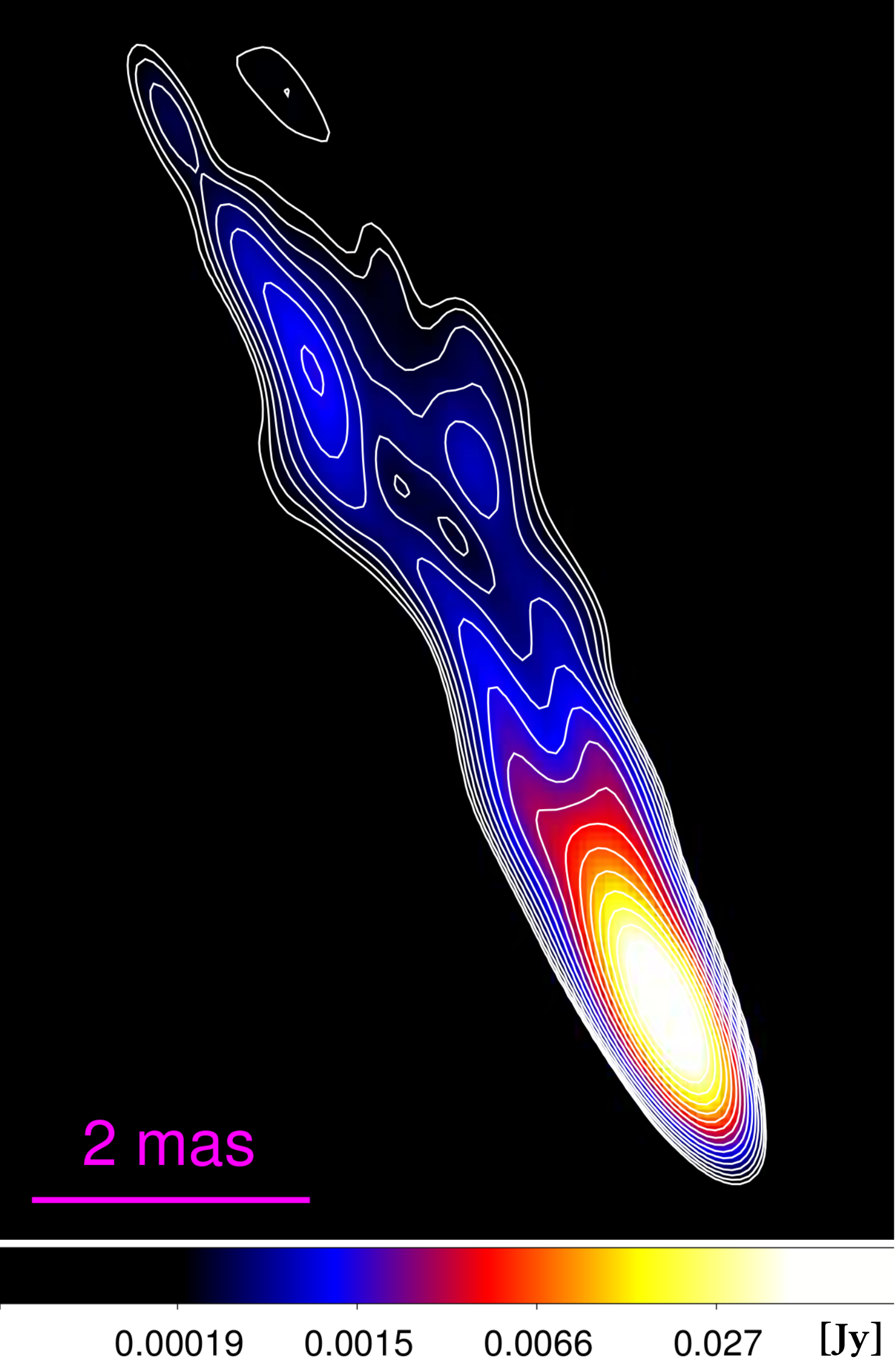}
\caption{Stacked VLBI image of 3C\,264 at 15 GHz. The jet emission appears limb-brightened. Contours represent isophotes at 0.17, 0.23, 0.31, 0.43, 0.61, 
0.89, 1.30, 1.92, 2.85, 4.24, 6.34, 9.48, 14.20, 21.29, 31.93, 47.90, 71.88, 107.88 mJy/beam. Each of the seven maps was convolved with a beam of $0.85\times 0.30$\,$\rm mas$, $25^{\circ}$.}
\label{stack}
\end{figure}

The radio data set comprises 7 VLBA (Very Long Baseline Array) observations at 15 GHz (2 cm) performed between September 2016 and October 2018 within the MOJAVE survey \citep{2018ApJS..234...12L}. The calibrated data were downloaded from the MOJAVE public archive, and the CLEAN images were produced in DIFMAP \citep[Differential Mapping,][]{1994BAAS...26..987S} through the {\sc CLEAN} algorithm after minor editing. In Table  \ref{VLBI} we report the log of observations and the characteristics of the seven maps. A stacked image, created after convolving each map with a common beam of $0.85\times 0.30$\,$\rm mas$, $25^{\circ}$, is shown in Fig. \ref{stack}. The beam position angle was set to coincide with the mean jet direction in order to more clearly show the internal structure of the flow, which is characterized by a prominent limb-brightening. The jet extends approximately for ${\sim}8$\,$\rm mas$, corresponding, at the source redshift and for the adopted cosmology, to ${\sim}3.5$ projected parsecs ($1\,\rm{mas}{\sim}0.44\,\rm{parsec}$). The faint jet emission appears smooth, with a single, main brightness enhancement (knot) at a radial distance of ${\sim}6$\, $\rm mas$. The radio core emission is moderately variable (Column 3 in Table \ref{VLBI}), showing an increase by approximately 20\% in 2018 with respect to the previous years.

\begin{table}
\centering
\caption{Log of VLBI observations at 15 GHz and characteristics of the CLEAN maps. Col. 1: Day of observation. Col. 2:  Beam FWHM and position angle. Col. 3: Peak flux density.  Col. 4: Total flux density. The uncertainties on the flux densities are of the order of $5\%$ \citep{2013AJ....146..120L}. Col. 5: Image noise. All values are given for untapered data with uniform weighting.}
\small
 \begin{tabular}{ccccc}
 \hline
 \hline
Date  & Beam & $S_{\rm peak}$ & $S_{\rm tot}$ & rms   \\
&{\footnotesize[$\rm mas\times \rm mas$, $\rm deg$]}&{\footnotesize[$\rm mJy/beam$]}&{\footnotesize[$\rm mJy$]}&{\footnotesize[$\rm \mu Jy/beam$]}\\
\hline
 26/09/16        & $0.80\times0.42, 2$     & 99    & 147         & 55                                       \\
 06/11/16        & $0.83\times0.41, -2$    & 101   & 143         & 55                                       \\
 10/12/16        & $1.09\times0.58, 22$    & 105   & 141         & 60                                       \\
 23/04/17        & $0.80\times0.39, 5$     & 97    & 148         & 85                                       \\
 30/07/17        & $0.87\times0.40, -5$    & 102   & 150         & 65                                       \\
 22/04/18        & $0.87\times0.43,  1$    & 119   & 171         & 50                                       \\
 06/10/18        & $0.82\times0.40, -3$    & 120   & 171         & 60                                       \\
 \hline
 \end{tabular}
 \label{VLBI}
\end{table}

\subsection{X-ray data}

3C\,264 was observed by the Chandra, XMM-Newton and Swift X-ray satellites (see Table~2). The XMM-Newton observation of 2001 was analyzed by \cite{2004ApJ...617..915D} while the Chandra observation of 2004 was presented by \cite{2006ApJ...642...96E}. In both analyses the source was characterized by a steep power law photon index $\Gamma{\sim}2.4$ and a luminosity of $\sim10^{42}$ erg s$^{-1}$ between 2-10 $\rm keV$. Thanks to an accurate analysis of the Chandra image from the 2004 observations, the X-ray counterpart of the radio-optical jet was later revealed by \cite{2010ApJ...708..171P}.

Here we present a comprehensive X-ray study  of the nuclear region of 3C\,264, collecting all the available data from the XMM-Newton, Chandra, and Swift/XRT archives. In three occasions, 3C\,264 was observed more than 8' off-axis, being the X-ray centroid of the cluster Abell\,1367 the primary target. On the contrary, Swift pointed directly at the source and performed 17 snapshots (less than 4 kiloseconds each) in 2018.

\begin{table*}
\caption{Nuclear emission: X-ray Spectral Parameters}
\small
\begin{center}
\begin{tabular}{lllll|lllll}
\hline
\hline 
Satellite  
&\multicolumn{1}{c}{Epoch} 
&MJD
&Exp$^{\alpha}$
& net counts
& $\Gamma$    
&Flux$^{\beta}$                   
& Lum$^{\zeta}$     
&NH$_{z}$$^{\star}$        
& $\chi^2$$^\S$/-Cstat$^{\dag}$/ \tabularnewline
& [year/month/day]
&    
& [sec]
&& &&&[dof]\tabularnewline  
 \hline
 &&&&&&&&\tabularnewline             
 Chandra               &  2000/02/26  & 51600      &  38950  & 8095  & 2.48$^{+0.08}_{-0.08}$  &  7.5$^{+0.6}_{-0.5}$   & 8      & 3$^{+1.0}_{-1.0}$    & 267/262  \tabularnewline
 XMM$^{\kappa}$        &  2001/05/26  & 52055      &  31740  & 3199  & 2.53$^{+0.08}_{-0.07}$  &  6.5$^{+0.5}_{-0.3}$   & 7      & $2\pm1$           & 293/270 \tabularnewline   
 Chandra              &  2004/01/24  & 53028      &  34760  & 10413 & 2.14$^{+0.06}_{-0.06}$            &  9.0$^{+0.6}_{-0.4}$   & 10     & $1.9^{+1.1}_{-1.0}$                 & 142/163 \tabularnewline
 XMM(MOS1)            &  2009/11/25  & 55160      &  10530  & 2079  & 2.35$^{+0.16}_{-0.15}$  &  14$^{+2}_{-1}$                  & 15     & $<4$          & 55/68    \tabularnewline  
 Swift                &  2018        &58132-58228 &  26570  & 2357  & 2.05$^{+0.12}_{-0.12}$  &  16$\pm1$                        & 17 &  $<3$                      & 100/99 \tabularnewline  
 &&&&&&&&\tabularnewline 
\hline
 &&&&&&&&\tabularnewline 
Swift    & 2018/01/14 & 58132 & 3341 &230   &  2.2$\pm0.2$       &  14$\pm2$         & 16  & ... &  9.3/16\\
Swift    & 2018/01/20 & 58138 & 2872 &233  &  2.2$\pm0.2$       &  13$\pm1$         & 15  & ... & 22/14\\
Swift    & 2018/01/24 & 58142 & 1386 &107  & 2.0$\pm0.3$        &  17$^{+4}_{-2}$ &  20 & ...  & 76c/83\\
Swift    & 2018/01/25 & 58143 & 2677  &248 & 1.9$\pm0.2$        &  10$\pm2$         & 13  &...   & 11/14\\
Swift    & 2018/03/18 & 58195 & 964 &121    & 2.0$\pm0.3$        &  19$^{+2}_{-4}$ &  25 &...   &  63c/95\\
Swift    & 2018/03/19 & 58196 & 2145 &254   & 2.1$\pm0.2$        &  23$\pm3$         &  30 &...   & 16/14\\
Swift    & 2018/03/20 & 58197 & 1503  &189 &1.9$\pm0.2$         &  22$^{+2}_{-4}$   &  27 &...   &  94c/133\\
Swift    & 2018/03/21 & 58198 & 1668&197   & 2.1$\pm0.2$        &  18$^{+2}_{-3}$ &  22 &...   &  137c/136\\
Swift    & 2018/03/23 & 58200 & 1838 &183   & 2.1$\pm0.2$        &  15$^{+2}_{-3}$ &  20 &...   &  90c/129\\
Swift    & 2018/03/24 & 58201 & 1376  &141 & 2.3$\pm0.2$        &  12$\pm2$         & 15  &...   &  77c/103\\
Swift    & 2018/04/08 & 58216 & 760 &57   & 2.1$\pm0.5$        &  12$\pm3$         & 16  &...   &  38c/47\\
Swift    & 2018/04/10 & 58218 & 1066 &116  & 2.1$\pm0.3$        &  18$^{+4}_{-2}$ &  21 &...   &  57c/94\\
Swift    & 2018/04/12 & 58220 & 899  &91    & 2.0$\pm0.3$        &  16$^{+2}_{-4}$ &  21 &...   &  57c/74\\
Swift    & 2018/04/14 & 58222 & 877  &90   & 2.2$\pm0.3$        &  11$^{+2}_{-3}$ &  15 &...   &  70c/68\\
Swift    & 2018/04/16 & 58224 & 967   &88  & 2.2$\pm0.3$        &  11$^{+2}_{-3}$ &  15 &...   &  61c/69\\
Swift    & 2018/04/18 & 58226 & 1024&77    & 2.0$\pm0.3$        &  12$^{+2}_{-3}$ &  15 &...   &  62c/62\\
Swift    & 2018/04/20 & 58228 & 1211 &121  & 1.8$\pm0.3$        &  20$^{+3}_{-5}$ &  26 &...   &  83c/93\\
&&&&&&&&   \tabularnewline  
\hline
 \multicolumn{10}{l}{$^{\alpha}$ -  Net exposure time.}\\
\multicolumn{10}{l}{$^{\beta}$ -  Flux in units of $10^{-13}$ erg cm$^{-2}$ s$^{-1}$ in the 2-10 keV range.}\\
\multicolumn{10}{l}{$^{\zeta}$ - Luminosity  in units of $10^{41}$ erg s$^{-1}$ in the 2-10 keV range.}\\
\multicolumn{10}{l}{$^{\kappa}$ - pn, MOS1 and MOS2 are fitted together using the same model but different power law normalisation.  Exposure time.}\\
\multicolumn{10}{l}{$^{}$ \,\,\,\, flux, and luminosity refer to MOS1.}\\
\multicolumn{10}{l}{$^{\star}$ - Column density in units of 10$^{20}$ cm$^{-2}$.}\\
\multicolumn{10}{l}{$^{\dag}$ -  {\it c} indicates C-statistic.}\\
\multicolumn{10}{l}{$^{\S}$ - When a thermal emission is included in the model, the $\chi^2$ value refers to the global fit.}\\
\end{tabular}
\end{center}
\vglue-0.1cm
\label{xcon}
\end{table*}

\begin{table*}
\caption{Thermal Emission: X-ray spectral parameters}
\small
\begin{center}
\begin{tabular}{lclll}
\hline
\hline 
Satellite  
& Date  
& kT  
&F$_{0.5-2~keV}$$^{\S}$
&L$_{0.5-2~keV}$$^{\S}$   \tabularnewline
                   
& 
&  [keV]  
& [erg s$^{-1}$ cm$^{-2}$] 
&[erg s$^{-1}$]  \tabularnewline

 \hline
  &&&&\tabularnewline             
Chandra           & 2000  & 0.6$^{+0.4}_{-0.3}$ & 3.6$(^{+3.1}_{-3.0}$)$\times10^{-14}$ &3.4$\times10^{40}$\tabularnewline
Chandra           & 2004  & 0.31$^{+0.11}_{-0.05}$ & 7.1($^{+2.4}_{-2.9}$)$\times10^{-14}$ &7.6$\times10^{40}$\tabularnewline
Swift/XRT         & 2018  & 0.2$^{+0.3}_{-0.1}$ & 12($^{+11}_{-10}$)$\times10^{-14}$ &13$\times10^{40}$\tabularnewline
&&&&\\
\hline
\multicolumn{5}{l}{$\S$ - Flux and luminosity are corrected for $N_{H}$ absorption.}
\end{tabular}
\end{center}
\vglue-0.1cm
\label{xtherm}
\end{table*}

All Chandra observations were performed using CCD/ACIS-S. The data were reprocessed by using CIAO version 4.10 with calibration database CALDB version 4.7.8, and applying the standard procedures. 
The radius $r$ of the extraction region depends on the distance of the source from the aim-points. We chose a circle of radius $r=5"$ when the source was on-axis and $r=27"$ for an off-axis position. This ensures the collection of at least 90$\%$ of counts in the selected regions \footnote{http://cxc.harvard.edu/proposer/POG/html/chap4.html}. The background spectra were extracted near the source in the same CCD.   Data were then grouped to a minimum of 15 counts per bin over the energy range $0.3-7$\,$\rm{keV}$. 

We checked if 3C\,264 was affected by severe pile-up in 2000 (when it was the main target of the pointing).
As a 1/8 sub-array configuration was adopted during the observation (time frame 0.41 s), we estimated\footnote{The tool PIMMSv4.9 was used to calculate the pile-up fraction: http://cxc.harvard.edu/toolkit/pimms.jsp} a negligible ($5\%$) pile-up. However, as a further check,  we also extracted a spectrum from an annulus  to exclude the inner region where a higher signal degradation is expected. The two spectra (circle and annulus) were fitted with the same model. The spectral parameters were completely consistent, although less statistically constrained in the annulus case.

In the two XMM-Newton observations, 3C\,264 was off-axis and therefore no pile-up effect is expected. These data were analysed using SAS version 14.0 and the latest calibration files. Periods of high particle background were screened by computing light curves above $10$\,$\rm keV$. In 2009, only the MOS1 spectrum could be analyzed. The pn camera was strongly affected by high background, while in MOS2 the source fell exactly in a CCD gap. Different radii were selected for the extraction circle of the source, depending on its position in the different cameras. In at least three cases, 3C\,264 was near the edge of a CCD and it was necessary to assume small regions. 
Spectra from the first observation (2001) were obtained by extracting events in circular regions of $r=40"$ (pn), $r=19"$ (MOS1), and $r=22"$ (MOS2).  For the only spectra available in 2009 a circle of radius $r=22"$ was adopted.
Data were then grouped to a minimum of 25 counts per bin over the energy range $0.3-7$\,$\rm keV$.  

Swift/XRT data were reduced using the data product generator of the UK Swift Science Data Center\footnote{http://www.swift.ac.uk/}.
%d
Source spectra for each observation were extracted from a circular region with a radius which depends on the count rate of the source. No pile-up effect reduced the quality of the data. The background was taken from an annulus with an inner radius of 142'' and outer radius of 260'' centered on the source. Spectra were grouped in order to have at least 1 count per bin \citep[see][for details]{2009MNRAS.397.1177E}. All Swift observations were also combined together in order to produce an average spectrum with higher signal to noise ratio. This could be grouped to a minimum of 15 counts in the 0.3-7 keV range allowing the application of the $\chi^2$-statistic.

Spectra were then fitted with XSPEC12.10 \citep{1999ascl.soft10005A}. The XMM spectra of the 2001 observation were fitted together using the same model. Only the flux was allowed to vary in order to take into account possible mis-calibration among the instruments. For Chandra and XMM-Newton, a $\chi^2$-statistic could be applied, while for the single Swift observations the Cstat-statistic was generally preferred, considering the short exposure and the limited number of counts. However, in at least 4 observations data could be grouped to a minimum of 15 counts per bin and the $\chi^2$-statistic adopted.  

We considered a power law model to fit the nuclear continuum and a thermal emission (APEC model) to reproduce the gas emission associated to the cluster Abell\,1367. The Galactic column density was fixed to $N_{H}$=1.82$\times10^{20}$\,$\rm cm^{-2}$ \citep{2005A&A...440..775K}. The presence of an intrinsic $N_{H,z}$, responsible for the attenuation of the nuclear emission, was also tested and required by all the Chandra and XMM-Newton observations. The $N_{H,z}$ intercepted by our line of sight is of a few 10$^{20}$\,$\rm cm^{-2}$, more probably related to the galaxy rather than to the nuclear region.  Hot emission due to the gaseous cluster environment was detected by Chandra and Swift/XRT (considering the average spectrum). The measured gas temperatures (kT${\sim}0.3$) and luminosities ($L_{0.5-2{\sim}\rm{keV}}\sim 10^{41}$\,$\rm{erg\cdot s^{-1}}$) are consistent within the uncertainties.  The large point spread function of the MOS off-axis and the short XRT integration time of each single observation prevent the detection of any low surface brightness extended emission. The results are reported in Tables \ref{xcon} and \ref{xtherm}. The parameter uncertainties correspond to 90\% confidence for a single parameter, i.e., $\Delta\chi^2=+2.7$ \citep{1976ApJ...210..642A}.

\begin{figure}[!h]
\centering  
\includegraphics[width=0.87\columnwidth]{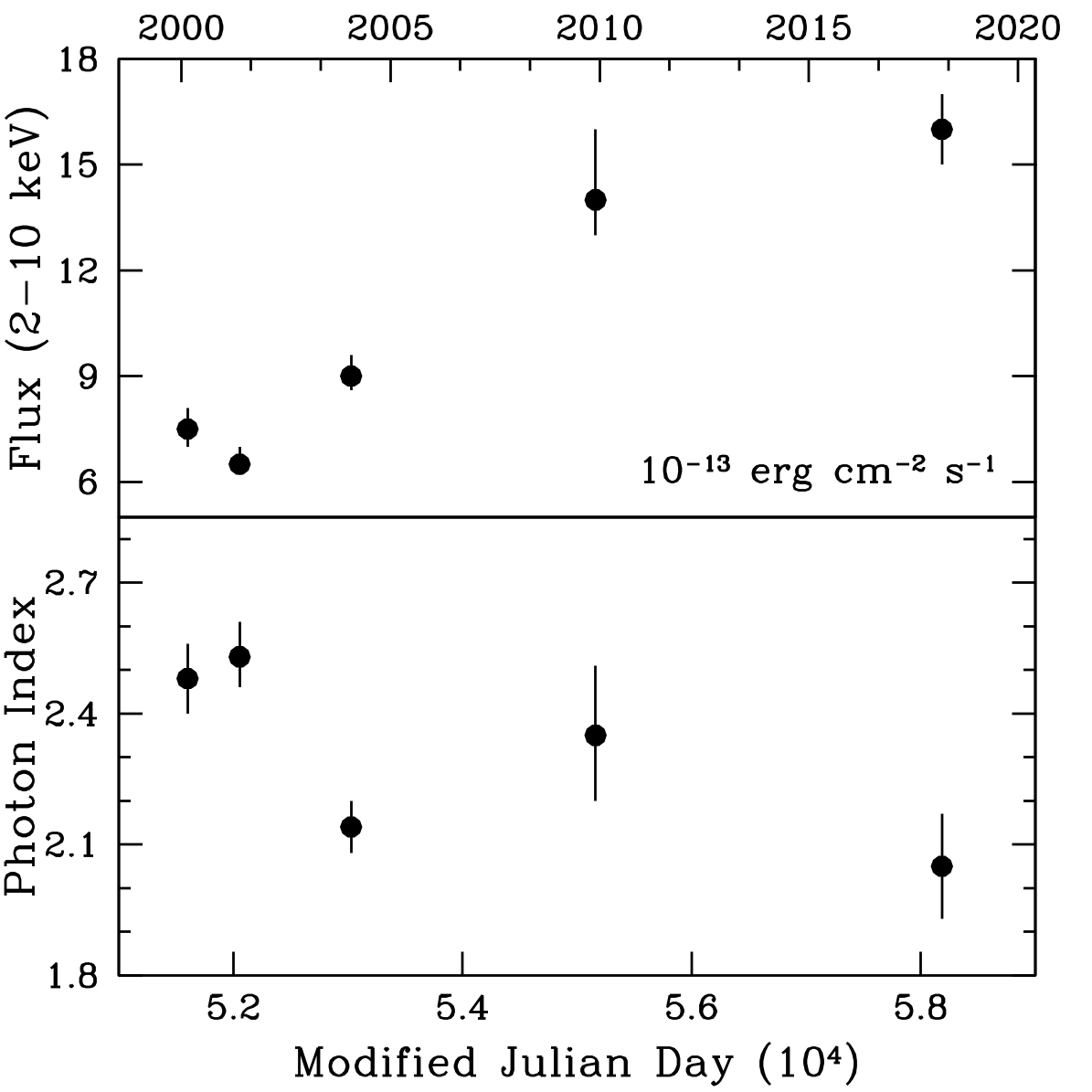}
\caption{Long-term variability from 2000 to 2018.  The nuclear (2-10 keV) flux increased by about a factor of 3 ({\it Upper panel}), while the nuclear emission became harder ({\it Lower Panel}).}
\label{lc}
\end{figure}

The light curve, shown in Figure \ref{lc} ({\it upper panel}), collects the $2-10$\,$\rm keV$ fluxes measured by all the satellites from 2000 to 2018. The Swift flux refers to an average value obtained from the collection of all the 2018 observations.  It is evident that the source luminosity changed by about a factor 3 from 2000 to 2018. In the {\it lower panel} the photon indices measured in the same period are shown. The investigation of a possible correlation between flux and photon index could, in principle, provide further insights on the physical mechanisms driving the variability. For instance, a ``hardening when brightening`` spectral behaviour is often observed in jets \citep[e.g.,][]{2017A&A...603A..25A}, and could be expected as a consequence of the injection of fresh energetic particles in the flow. However, due to the bias introduced by integrating the spectrum to calculate the flux \citep[see][]{1996A&A...312..810M}, the present data are insufficient to attest the statistical significance of a correlation between $\Gamma$ and flux. This may be better explored in the future based on a denser monitoring of the source.

\begin{figure}[!h]
\centering   
\includegraphics[width=0.94\columnwidth]{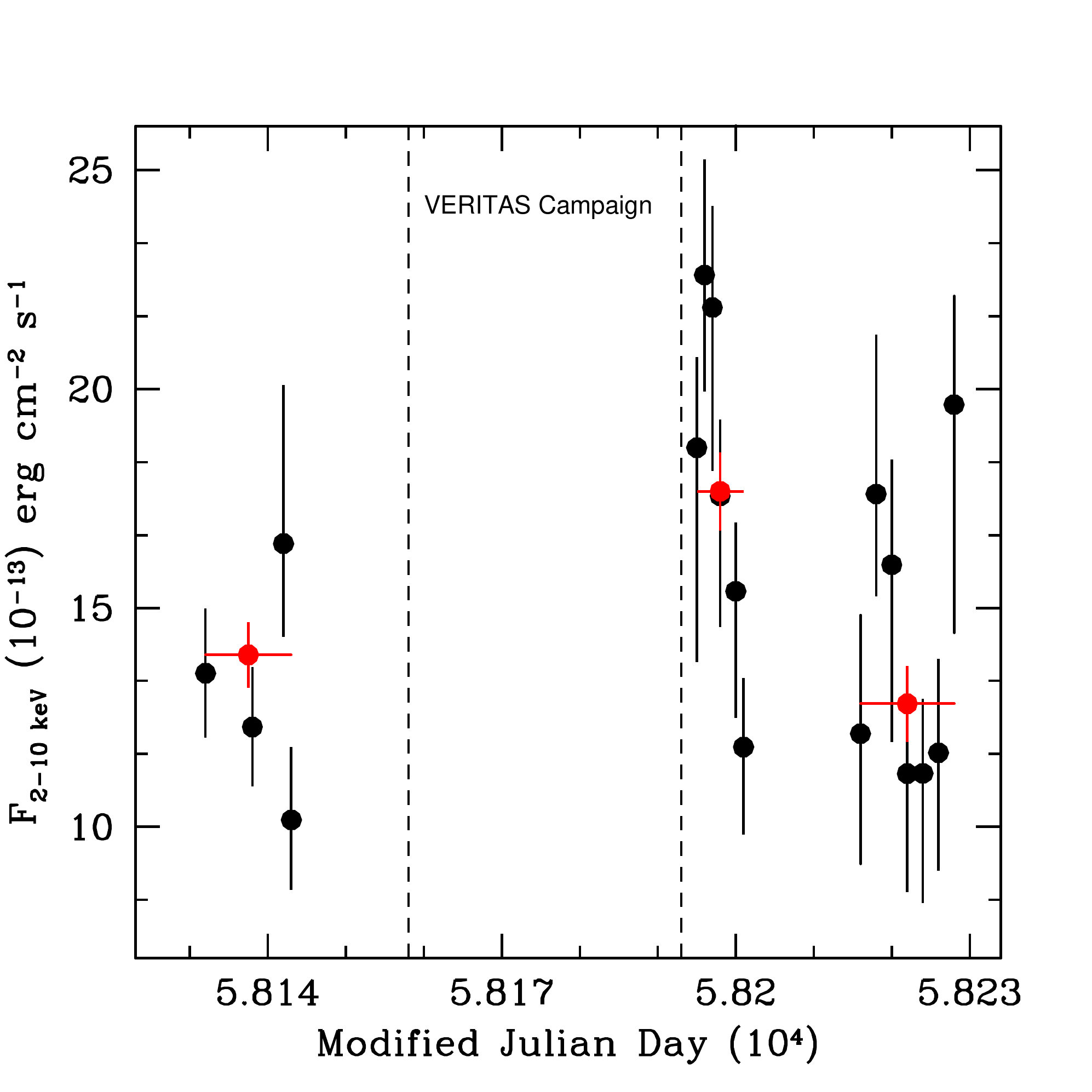}
\caption{Short-term flux variability revealed by Swift. Black points correspond to the single observations performed in 2018 from January to April. Red points represent the fluxes averaged on each single month (1-$\sigma$ error-bars).  The source appears to be in a high state in March, two days after the end of the VERITAS Campaign.}
\label{lcs}
\end{figure}

Flux variations are also observed on shorter time scales. Figure \ref{lcs} shows the Swift light curve from January 2018 to April 2018. In spite of the large uncertainties, there is an indication for a flare episode in March 2018. In order to statistically test the high state of the source in that period, all the observations performed in a single month were collected together. Each spectra could be grouped in order to have at least 20 counts per bin and the chi-statistic applied. A power law absorbed by Galactic column density was a good fit in all the three cases (see Table \ref{xcon}). The average January, March, April fluxes are shown in Figure \ref{lcs} as red points, with error-bars corresponding to 1-$\sigma$ (68\% confidence level for two parameters of interest i.e. $\Delta\chi^2$=2.3). A standard $\chi^2$ test, applied to the average fluxes (red points in Figure \ref{lcs}), confirms the variability of 3C\,264, being  the probability to be constant less than $10^{-3}$.
Interestingly, the X-ray enhancement was observed two days after the end of the VERITAS campaign that detected for the first time 3C\,264 in the TeV band. The two bands could then be related to each other and the TeV photons produced by up-scattering of keV photons in the jet. 

\subsection{$\gamma$-ray data}

In the $\rm GeV$ band, 3C\,264 is associated with 3FGL J1145.1+1935, a source reported in the {\it Third Fermi Large Area Telescope Source Catalog} \citep{2015ApJS..218...23A}. It was detected at the 5.7$\sigma$ significance level with a flux in the 1-100 $\rm GeV$ band of $(2.6\pm 0.7)\times 10^{-10}$\,$\rm ph\cdot cm^{-2}\cdot s^{-1}$ and a power law photon index $\Gamma_{3FGL}=$1.98$\pm$0.2.
The $\rm GeV$ spectrum of 3C\,264 is quite hard compared to the other radio galaxies with a $\gamma$-ray counterpart. Indeed, the source is also listed in the third Fermi-LAT catalog of hard-spectrum sources \citep{2017ApJS..232...18A}, containing all objects detected in seven years of survey above 10\,$\rm GeV$. In the hard regime, its power law index is $\Gamma_{>10GeV}=1.65\pm0.33$ and the flux between 10\,$\rm GeV$ and 1\,$\rm TeV$ $F_{3FHL}=(2.5\pm 1.0)\times 10^{-11}$\,$\rm ph\cdot cm^{-2}\cdot s^{-1}$. No significant flux variation on time scales of one month is reported for this source in the 3FGL catalog.

\begin{figure*}
\centering
\includegraphics[trim=0cm 0cm 0cm 0cm, clip=true, width=0.225\textwidth]{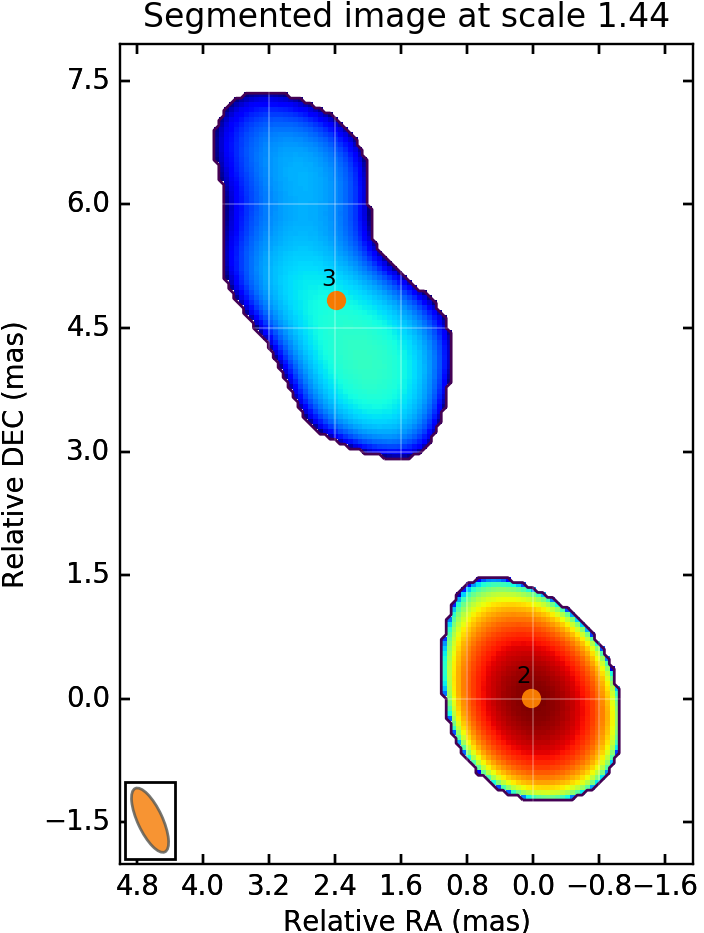}
\includegraphics[trim=0cm 0cm 0cm 0cm, clip=true, width=0.225\textwidth]{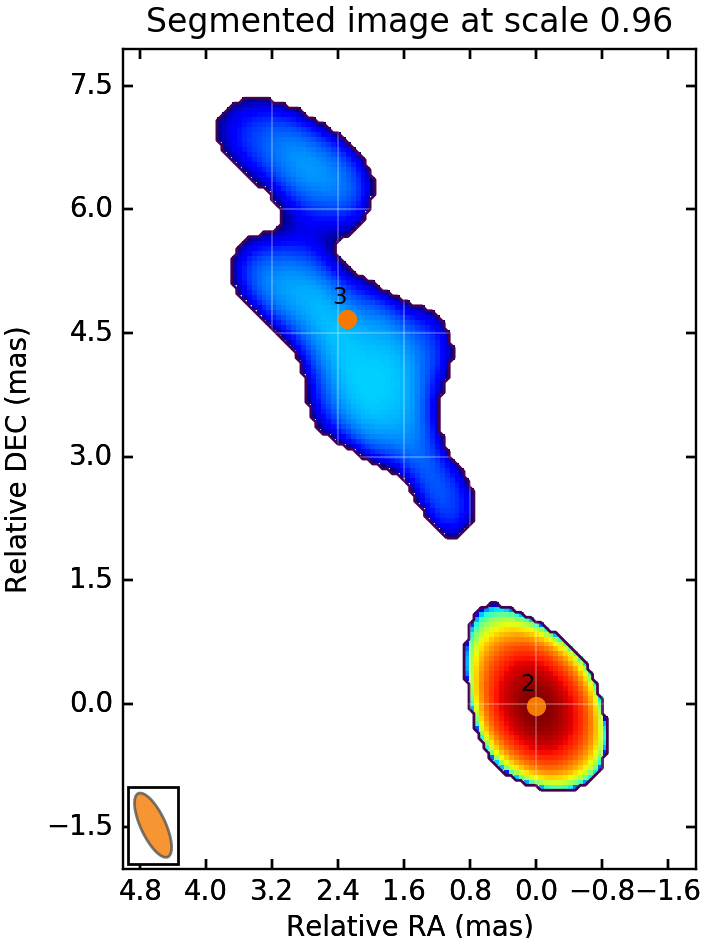}
\includegraphics[trim=0cm 0cm 0cm 0cm, clip=true, width=0.225\textwidth]{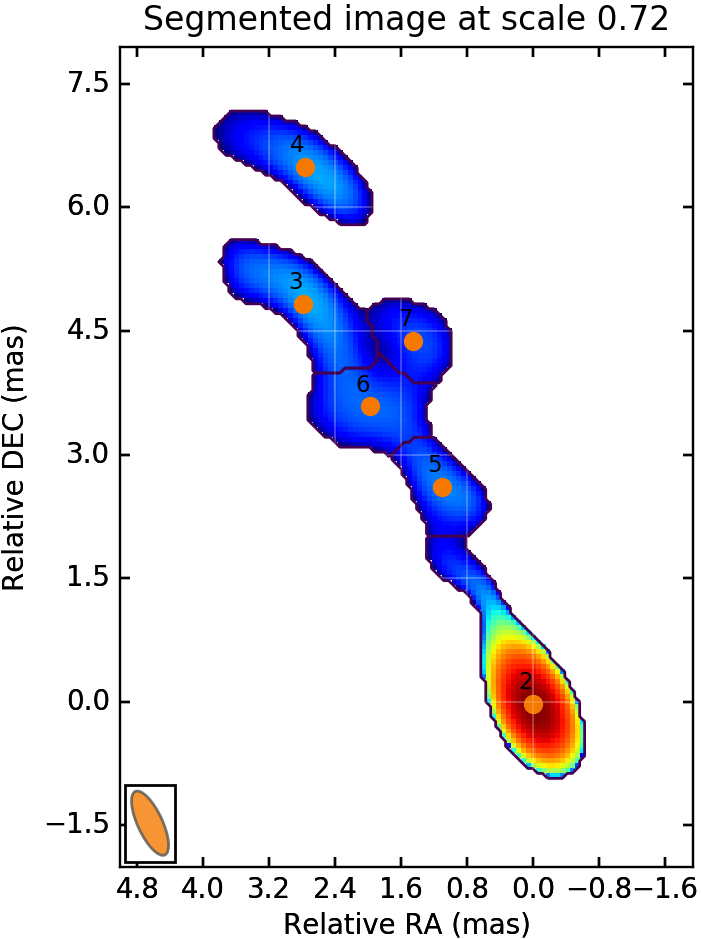}
\includegraphics[trim=0cm 0cm 0cm 0cm, clip=true, width=0.225\textwidth]{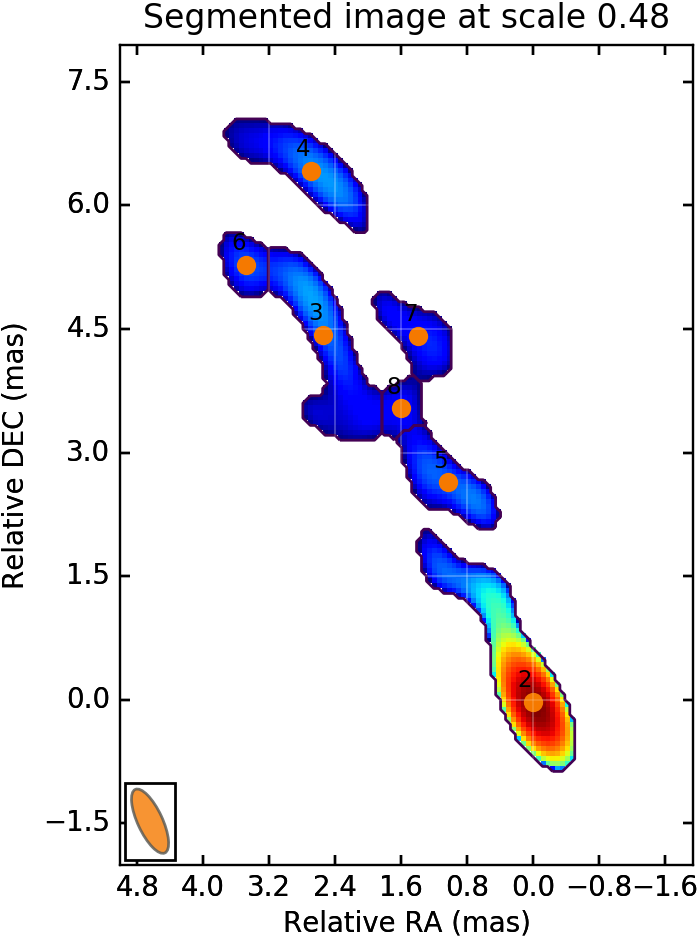}\\
\vspace{0.134cm}
\includegraphics[trim=0cm 0cm 0cm 0cm, clip=true, width=0.225\textwidth]{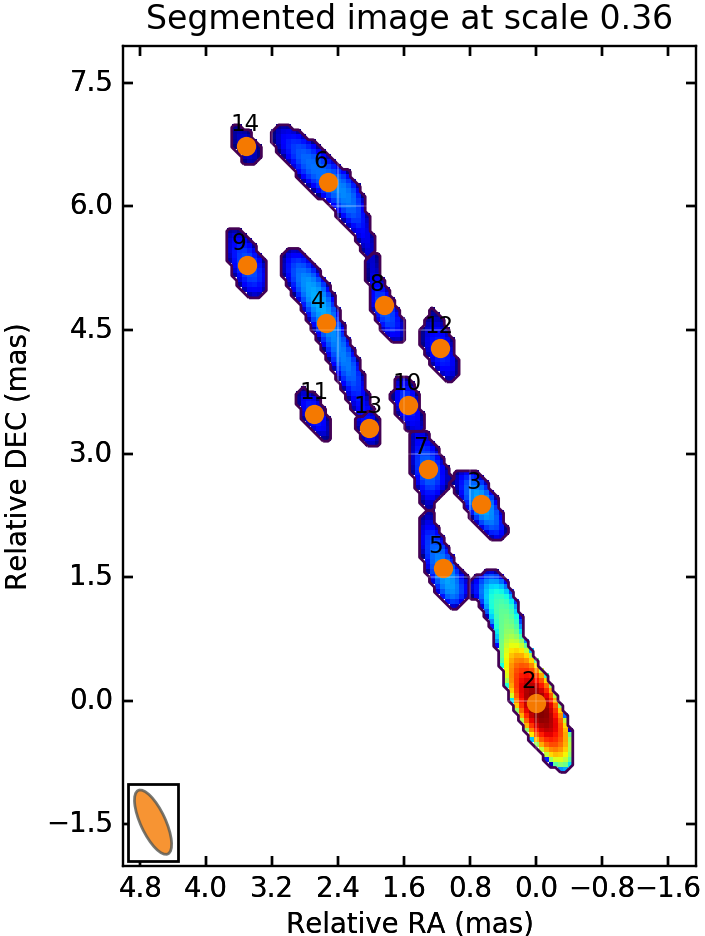}
\includegraphics[trim=0cm 0cm 0cm 0cm, clip=true, width=0.225\textwidth]{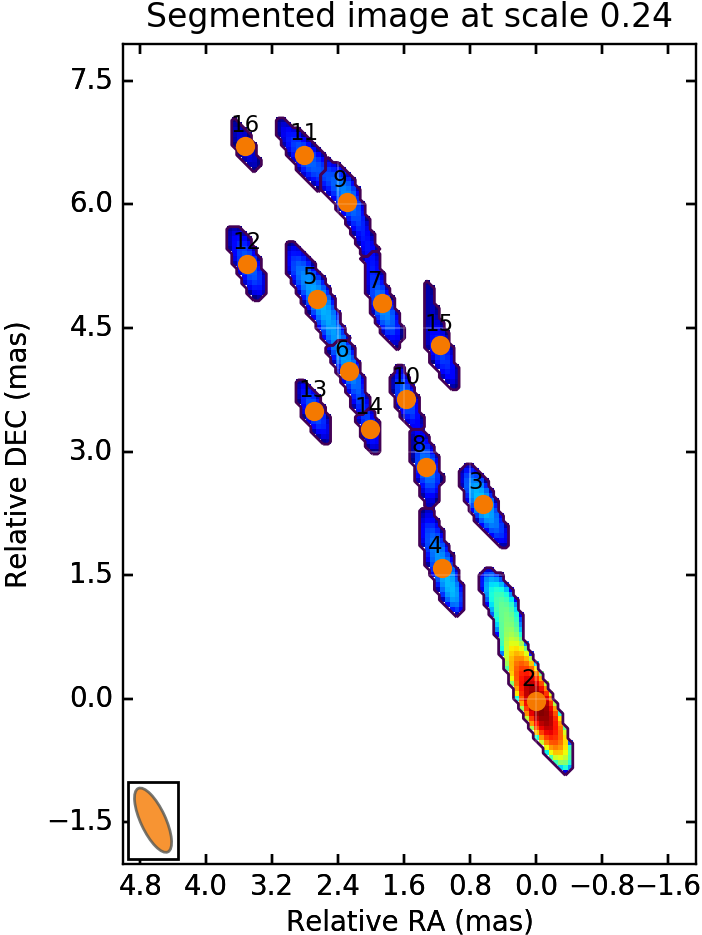}
\includegraphics[trim=0cm 0cm 0cm 0cm, clip=true, width=0.225\textwidth]{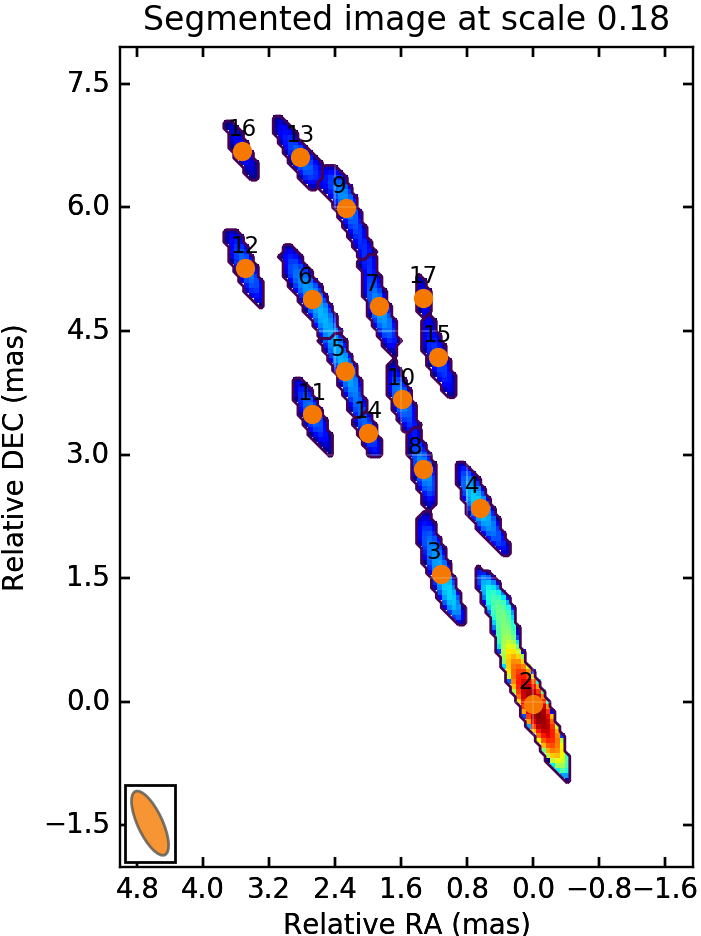}
\includegraphics[trim=0cm 0cm 0cm 0cm, clip=true, width=0.225\textwidth]{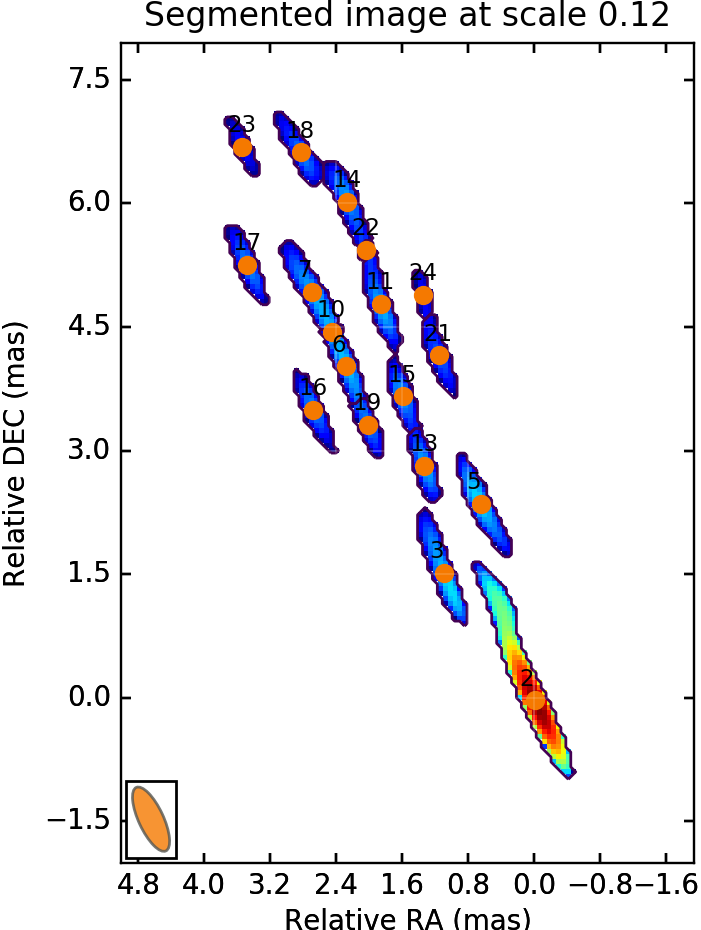}
\caption{Multi-scale wavelet decomposition of the jet in 3C\,264, as observed in July 2017. Orange points mark the position of each jet feature. Features are numbered in order of date (oldest first) and flux density (brightest first). The scales are expressed in units of milli-arcseconds.  }
\label{July}
 \end{figure*}
 \vspace{0.2cm}
 \begin{figure*}
 \centering
\includegraphics[trim=0cm 0cm 0cm 0cm, clip=true, width=0.225\textwidth]{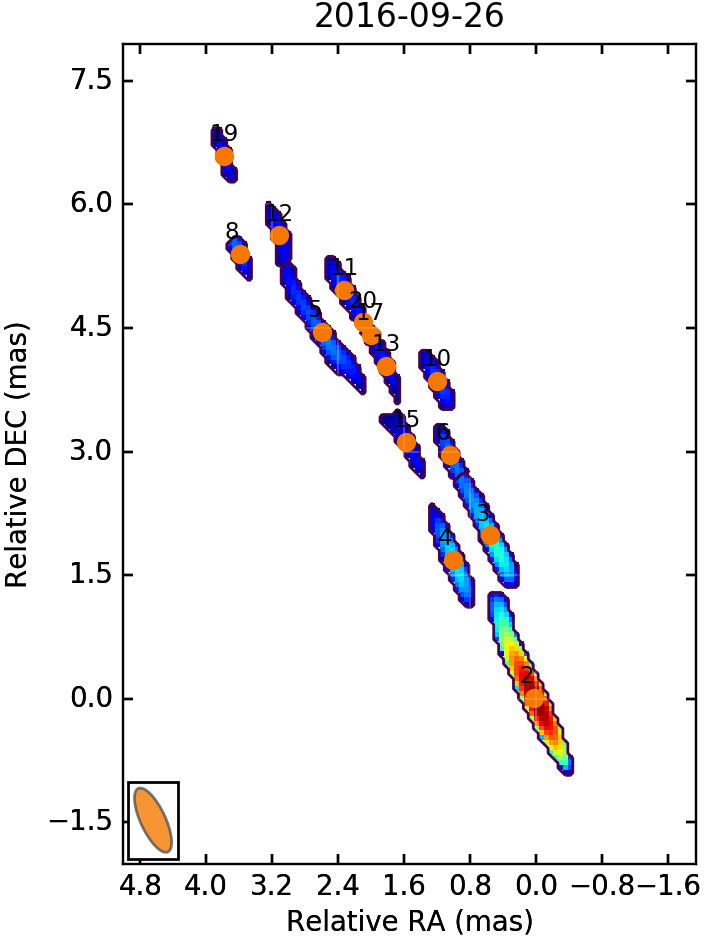}
\includegraphics[trim=0cm 0cm 0cm 0cm, clip=true, width=0.225\textwidth]{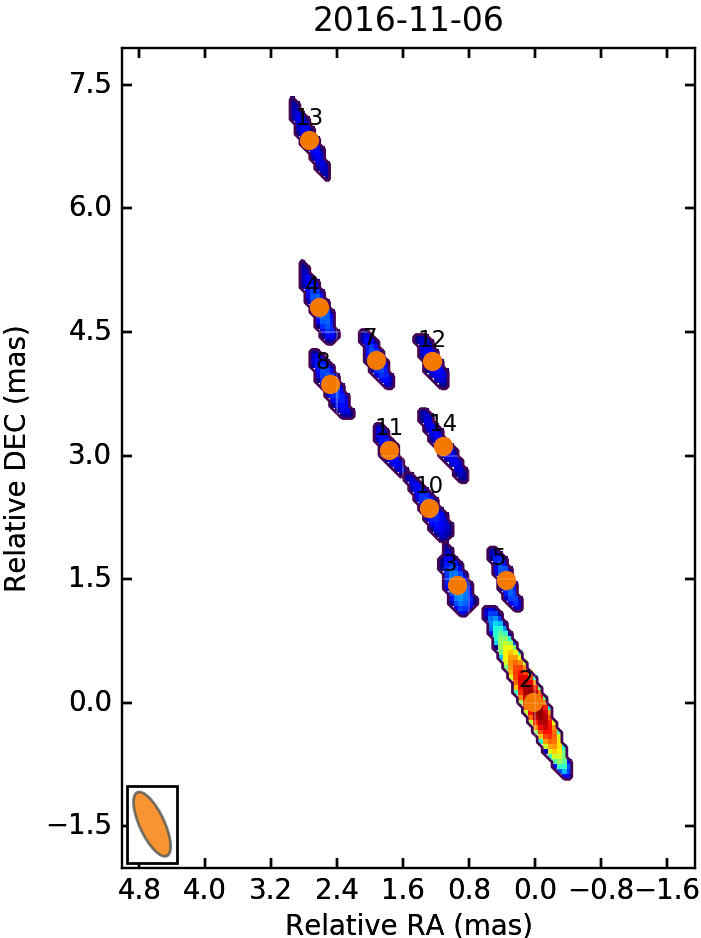}
\includegraphics[trim=0cm 0cm 0cm 0cm, clip=true, width=0.225\textwidth]{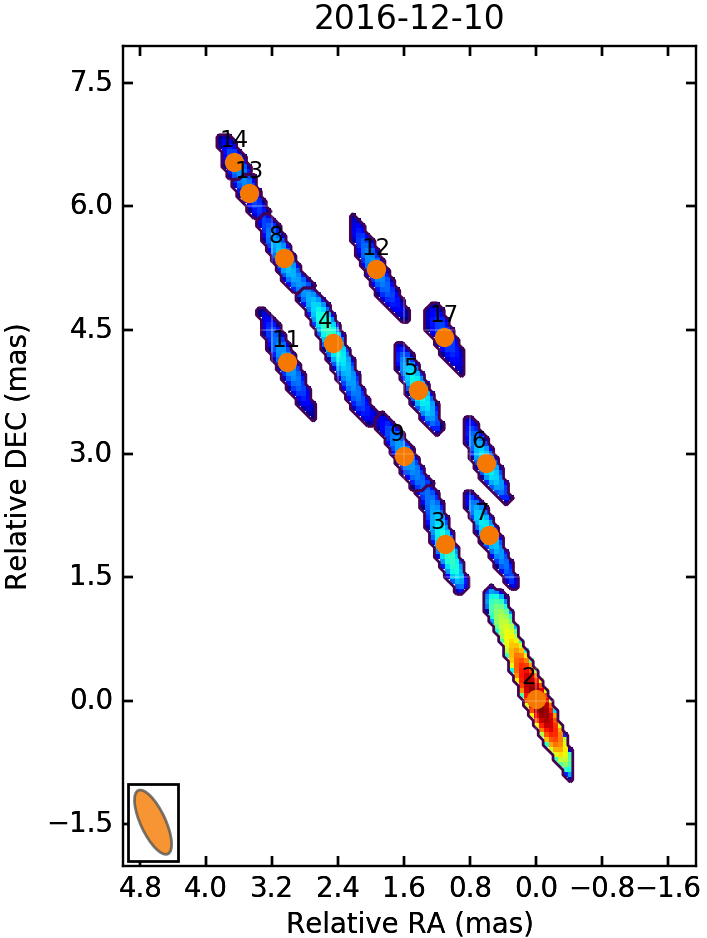}
\includegraphics[trim=0cm 0cm 0cm 0cm, clip=true, width=0.225\textwidth]{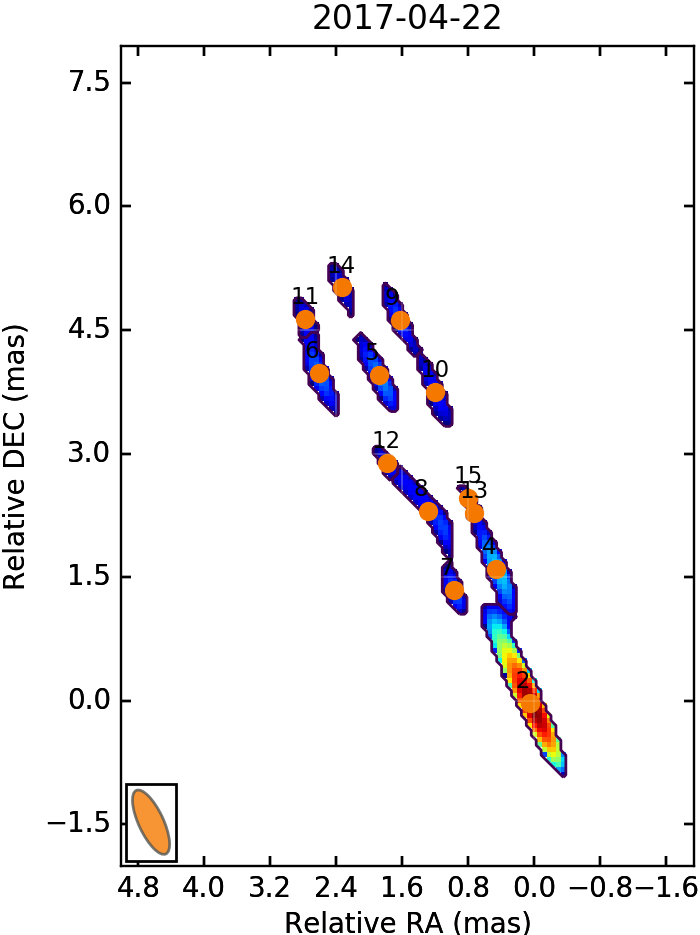}\\
\vspace{0.134cm}
\includegraphics[trim=0cm 0cm 0cm 0cm, clip=true, width=0.225\textwidth]{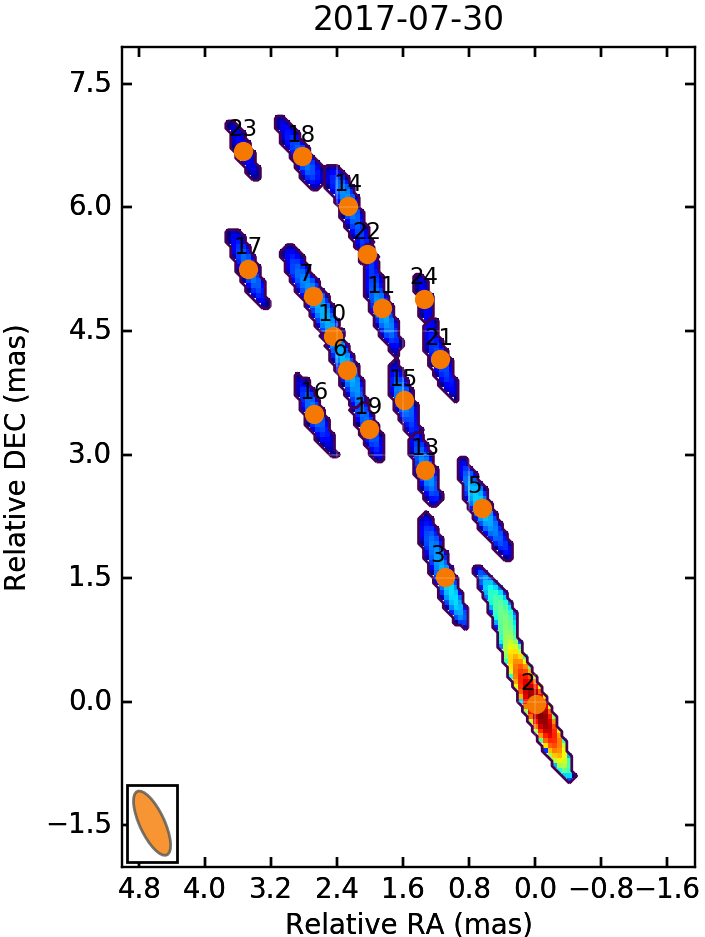}
\includegraphics[trim=0cm 0cm 0cm 0cm, clip=true, width=0.225\textwidth]{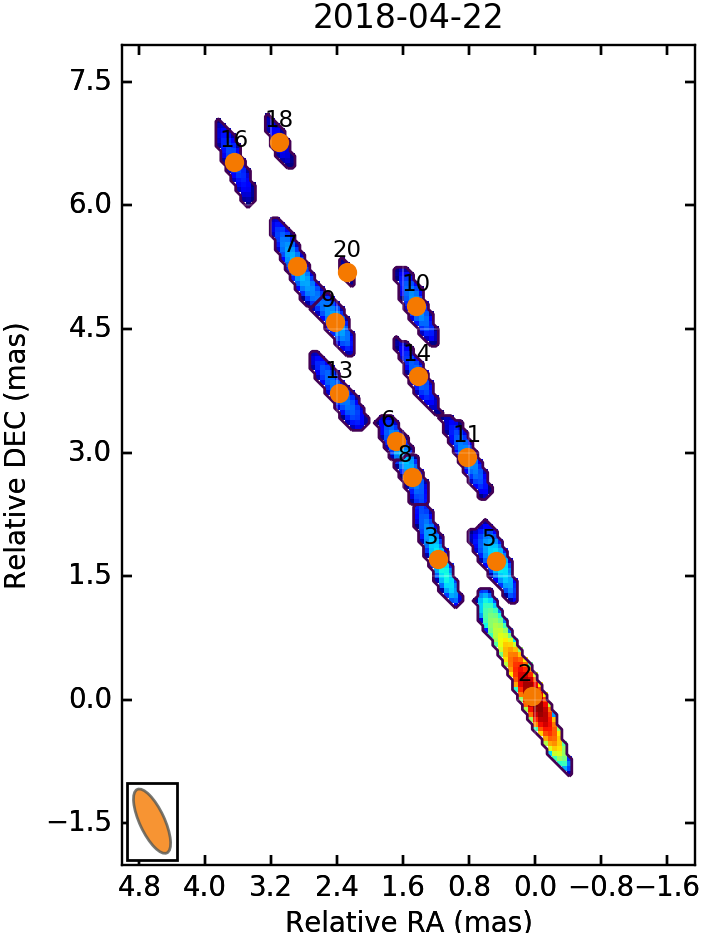}
\includegraphics[trim=0cm 0cm 0cm 0cm, clip=true, width=0.225\textwidth]{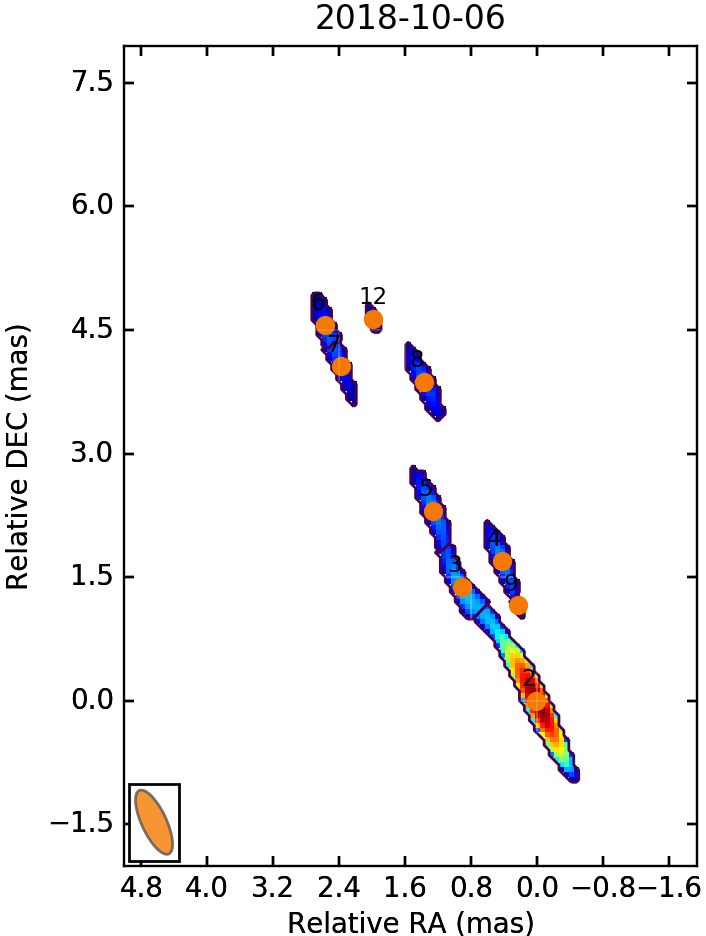}
\caption{Wavelet decomposition at the spatial scale of 0.12 mas, for all epochs. Due to variations of the local signal-to-noise ratio, only some parts of the jet filamentary structure are visible at each time.}
\label{s2}
 \end{figure*}
\section{Results from VLBI}
\subsection{Kinematics}
The kinematic analysis of the radio jet was performed using the the publicly available Wavelet Image Segmentation and Evaluation (WISE) code \citep{2015A&A...574A..67M, 2016A&A...587A..52M}. By exploiting a segmented wavelet decomposition method (SWD), the WISE code enables a multi-scale detection of statistically significant structural patterns (SSP) in the examined astronomical image. Given a multi-epoch set of images, the significant patterns are then tracked in time by running a multi-scale cross-correlation algorithm (MCC) between pairs of consecutive epochs, which yields a measurement of the pattern displacement. 

The chosen approach is ideal for the kinematic study of the jet in 3C\,264, which is transversely resolved and lacks bright and well defined knots. Traditional methods, like the MODELFIT subroutine in DIFMAP \citep{1994BAAS...26..987S}, rely on a priori assumptions concerning the shape of the structures contributing to the brightness distribution. For instance, in the case of jets, a set of two-dimensional Gaussian components is typically fitted to the data. The wavelet decomposition, on the other hand, enables arbitrarily shaped structures to be identified. 

The maps considered in the SWD analysis were slightly super-resolved in the direction perpendicular to the jet axis, being restored with a common beam of $0.85\times0.30$\,$\rm mas$, $25^{\circ}$ (the same used to create the stacked image in Fig.\ref{stack}). These parameters were chosen to facilitate the investigation of the jet kinematic structure in the transverse direction. To cross-check the results obtained, the analysis was also performed after restoring all images with the average natural beam, which is approximately $0.85\times0.40$\,$\rm mas$, $5^{\circ}$.

\subsubsection{Image decomposition}

The segmented wavelet decomposition was carried out for each epoch on a set of five spatial scales (0.12 mas, 0.24 mas, 0.48 mas, 0.72 mas, 1.44 mas), plus three intermediate scales (0.18 mas, 0.36 mas, 0.96 mas). The intermediate wavelet decomposition (IWD), described in \cite{2016A&A...587A..52M}, is aimed at recovering structural information in optically thin, stratified flows, where multiple patterns can overlap without inducing full obscuration. The complex limb-brightened jet structure shown in Fig. \ref{stack} suggests that this is most likely the case in 3C\,264. 

 In Figure \ref{July} we present an example of the multi-scale wavelet decomposition of the jet in 3C\,264, using data from July 2017. While at the largest spatial scale of 1.44 mas the algorithm detects only 2 components, the smaller scale decomposition obviously yields an increasingly rich set of information, with the detection of 18 components at the smallest spatial scale of 0.12 mas. In the specific observation presented in Fig. \ref{July} the jet shows a quite complex structure, being composed by four filaments which appear to possibly intersect and cross each other. Such pattern is however not clearly seen in other maps. 
 As shown in Fig. \ref{s2}, a smaller number of filaments and/or features is detected in other epochs. This is most likely due to variations of the local signal to noise ratio with time, with only some portions of the jet being illuminated in a single epoch.

\subsubsection{Cross-identification and proper motions}
Once the significant patterns are identified, the multi-scale cross-correlation algorithm matches them by examining pairs of adjacent epochs. The matching of features detected on the larger scales is used as a reference for the robust cross-identification of features on the smaller scales. In the MCC analysis, we assumed a correlation threshold of 0.6 and we enabled the displacements to vary between 0 and 12 $\rm mas/year$ in the longitudinal direction and between -6 and 6 $\rm mas/year$ transversely. About 59\% of the total features detected at all scales was successfully cross-identified, with the normalized cross-correlation coefficient being systematically larger than 0.75. When considering only the results obtained at the smallest scale of 0.12 mas, the success rate of the matching algorithm decreases to 39\%. 

 \begin{figure}[!h]
\centering 
\includegraphics[trim=0cm 0cm 0cm 0cm, clip=true, width=0.49\textwidth]{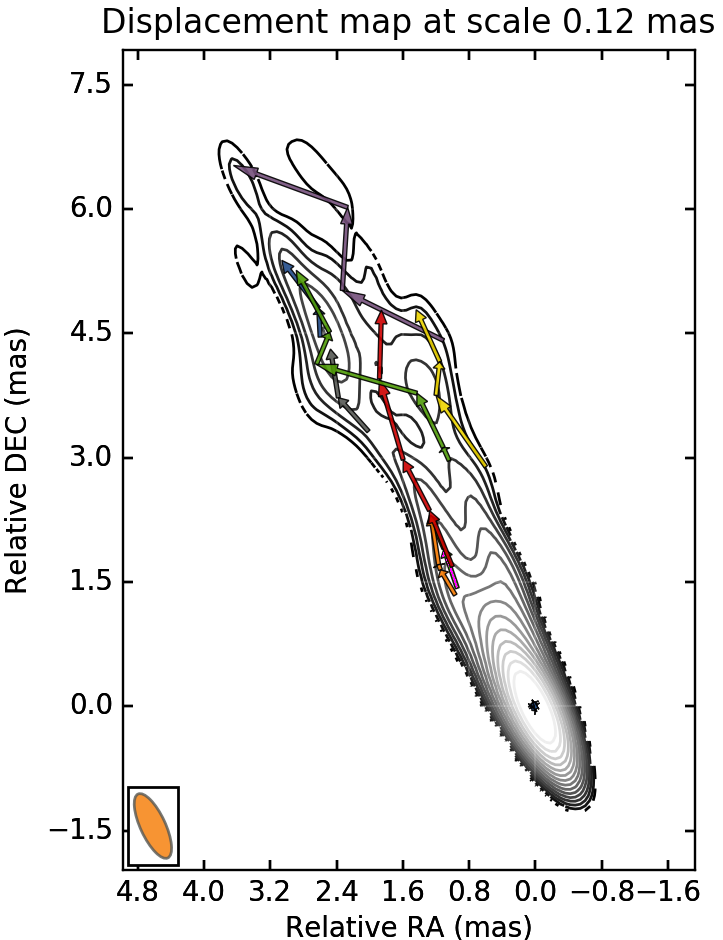}
\caption{Displacement vectors of all features matched in at least three epochs at the finest wavelet scale of 0.12 mas. The vectors are overlaid on the stacked image. Each feature is marked by a different color.}
\label{vec}
\end{figure}

In Figure \ref{vec} we present the displacement vectors of the patterns that were successfully cross-identified in at least three epochs, overlaid on the stacked image. While we show the results obtained at the finest scale, we remark that the displacements determined at larger scales are very consistent. 
The features are observed to move on complex trajectories, mainly following the jet direction but also showing a significant transverse component in some cases. In the inner jet, up to a distance of $\sim3$\,$\rm mas$, most of the features that could be matched belong to the eastern limb. In the outer jet, instead, the proper motion is detected on both limbs. In addition, thanks to the fact that the jet transverse structure is better resolved at larger distances, in the outer jet we can have a closer look at kinematic properties near the jet axis. The displacement of features detected in the central body of the jet is characterized by a larger transverse component. In particular, the jet limbs appear to clearly intersect and cross each other at a projected distance from the core of $\sim4.5$\,$\rm mas$, in the proximity of the brightest jet knot. 

\begin{figure}[!h]
\centering 
\includegraphics[trim=0cm 0cm 0cm 0cm, clip=true, width=0.5\textwidth]{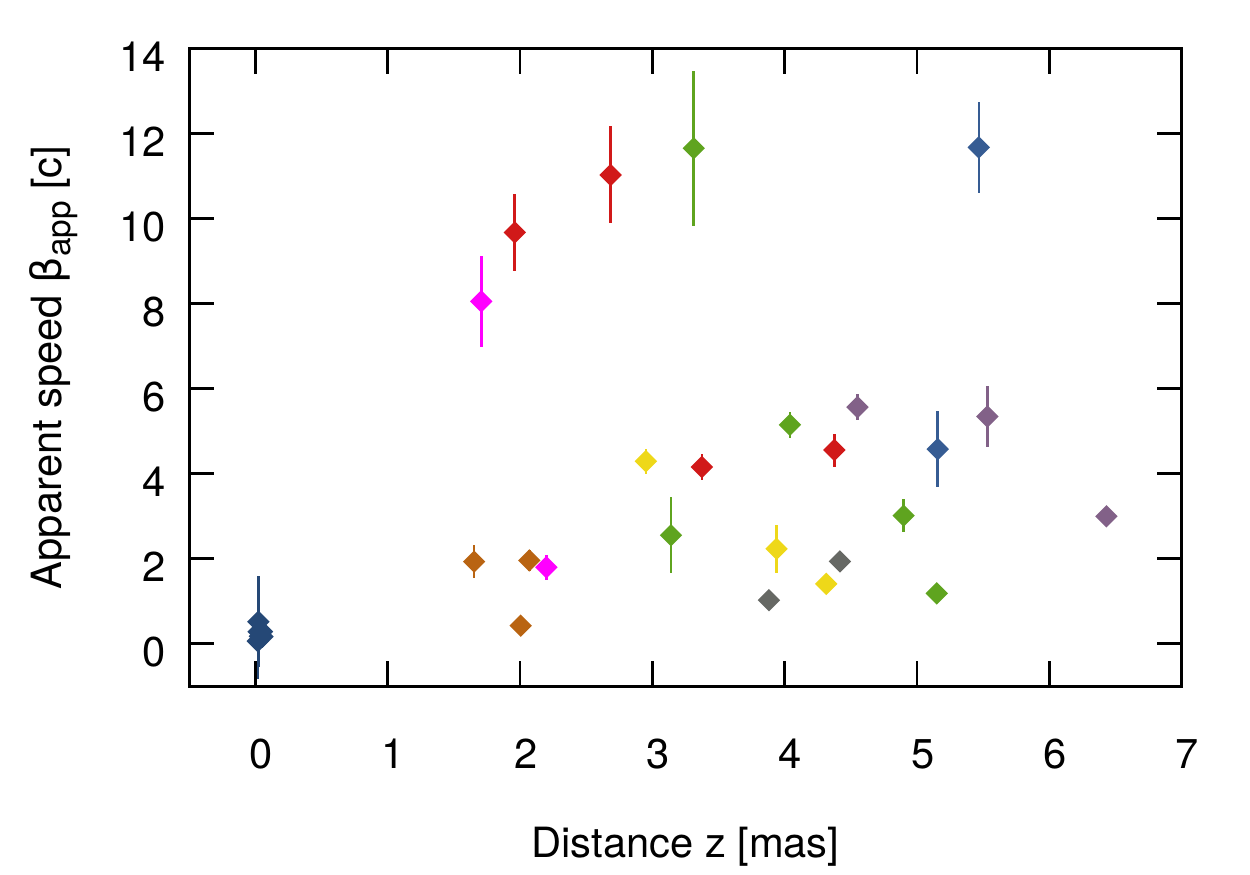}
\caption{Apparent velocities as a function of radial distance from the core. Each feature is marked by a different color, coded in agreement with Fig. 6.}
\label{speed}
\end{figure}

From the proper motions, we have then computed the apparent velocity $\beta_{\rm app}$ of each feature in units of the speed of light $\rm c$. At the source redshift and for the adopted cosmology, a proper motion of $1\,\rm{mas/year}$ corresponds to an apparent speed of $1.52\,c$.\\ The uncertainties on the velocities are calculated as described in \cite{2016A&A...587A..52M} by accounting for the positional error of each SSP, which in turn depends on the signal-to-noise ratio of the SSP detection. In Figure \ref{speed}, we report the obtained values of $\beta_{\rm app}$ as a function of radial distance from the jet core $z$. In this plot, the color assigned to each feature matches the color of the displacement vectors shown in Fig. \ref{vec}.
The velocities are distributed in a broad interval between ${\sim}0.5\,\rm{c}$ and ${\sim}11.5\,\rm{c}$. The presence of a gap is apparent between a lower envelope of speeds reaching a maximum value of ${\sim}5\,\rm{c}$, and an upper envelope peaking at ${\sim}11\,\rm{c}$. Both the upper and the lower envelope show a trend of increasing velocity with distance up to ${\sim}3.5-4.5$\,$\rm{mas}$ (or ${\sim}1.5-2$ projected parsecs) from the core, followed by a possible flattening at larger separation. Similar acceleration profiles have been recently obtained in the kinematic studies of M\,87 \citep{2016A&A...595A..54M} and Cygnus\,A \citep{2016A&A...585A..33B}, in agreement with the theoretical prediction that magnetic acceleration of the bulk flow proceeds slowly and extends on parsec scales \citep[e.g.,][]{2004ApJ...605..656V}.

\begin{figure}[!h]
\centering 
\includegraphics[trim=0cm 0cm 0cm 0cm, clip=true, width=0.5\textwidth]{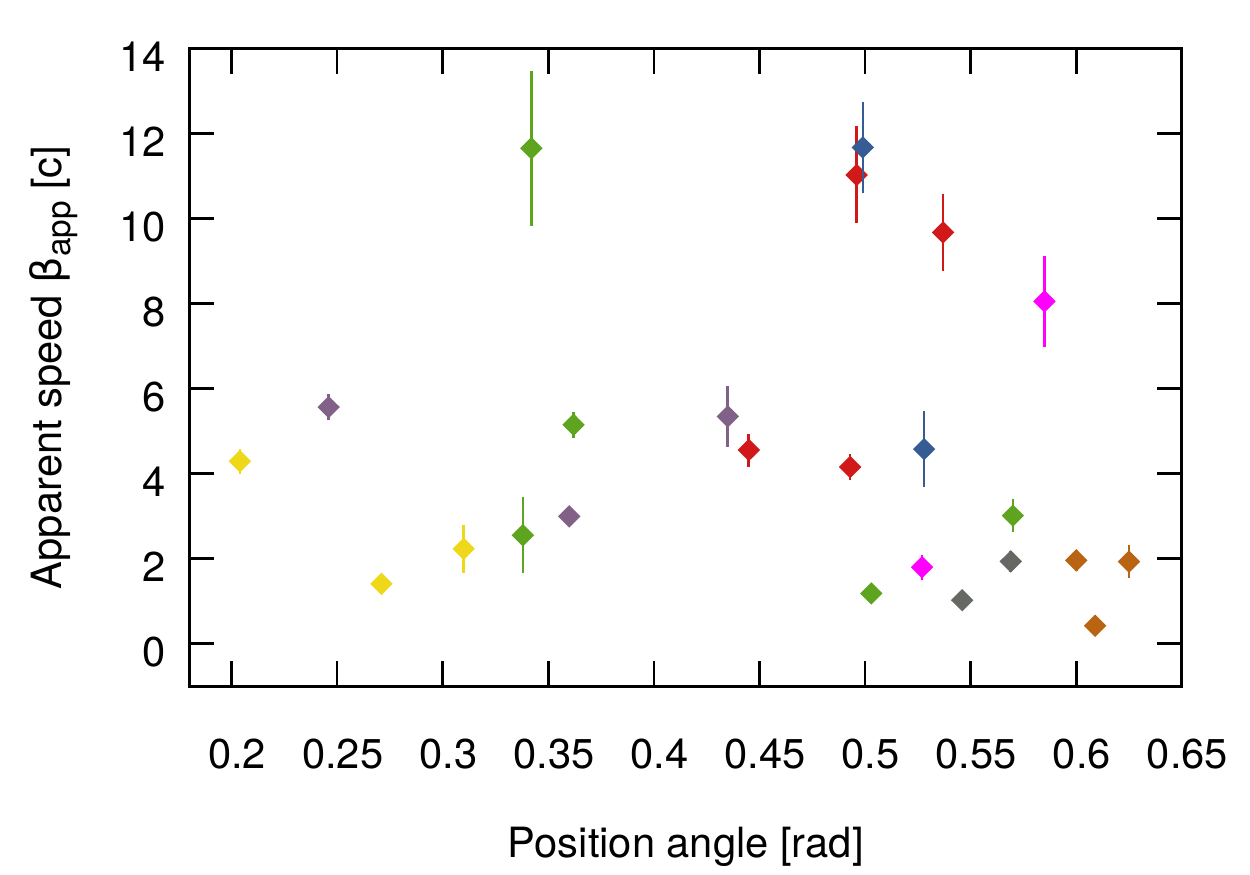}
\caption{Apparent velocities as a function of position angle. The jet axis is oriented at an average position angle of ${\sim}0.44$\,$\rm radiants$. Each feature is marked by a different color, coded in agreement with Fig. 6 and Fig. 7.}
\label{pos}
\end{figure}
  
In Figure \ref{pos}, we also examine how the apparent velocities are distributed in the transverse direction. To this aim, we plot the apparent speeds as a function of position angle of the features. The jet axis is oriented at an average position angle of ${\sim}0.44$\,$\rm radiants$. The SSP are detected in a range between ${\sim}0.2$ and ${\sim}0.65$\,$\rm radians$ and, as already discussed above, they tend to cluster at the edges of this interval, i.e. on the jet limbs. The high velocities, above ${\sim}5$\,$\rm c$, are found either close to the jet axis or along the eastern limb. Therefore from our analysis we do not infer a clear speed stratification, e.g., of the spine-sheath kind. It is possible, however, that such a velocity gradient does exist in the flow, but the low and the high speed components appear mixed due to the overlap of optically thin filaments seen in projection. This scenario was proposed previously for the interpretation of the kinematic properties of the M\,87 jet \citep{2016A&A...595A..54M}, where a spine-sheath structure is also not directly and unambiguously inferred from the kinematic study, even though the plasma flow appears prominently stratified.  

As mentioned in Sect. 4.1, the SWD analysis was performed a second time by considering maps convolved with the average natural beam, which is approximately $0.85\times0.40$\,$\rm mas$, $5^{\circ}$, with the aim of cross-checking the robustness of the kinematic results.  While the algorithm cannot successfully detect and/or match all of the features in Fig. \ref{speed} due to the lower resolution in the transverse direction, both the displacement map and the range of apparent speeds inferred though this second approach are very consistent. 

\begin{figure*}
    \centering
    \includegraphics[width=0.9\textwidth]{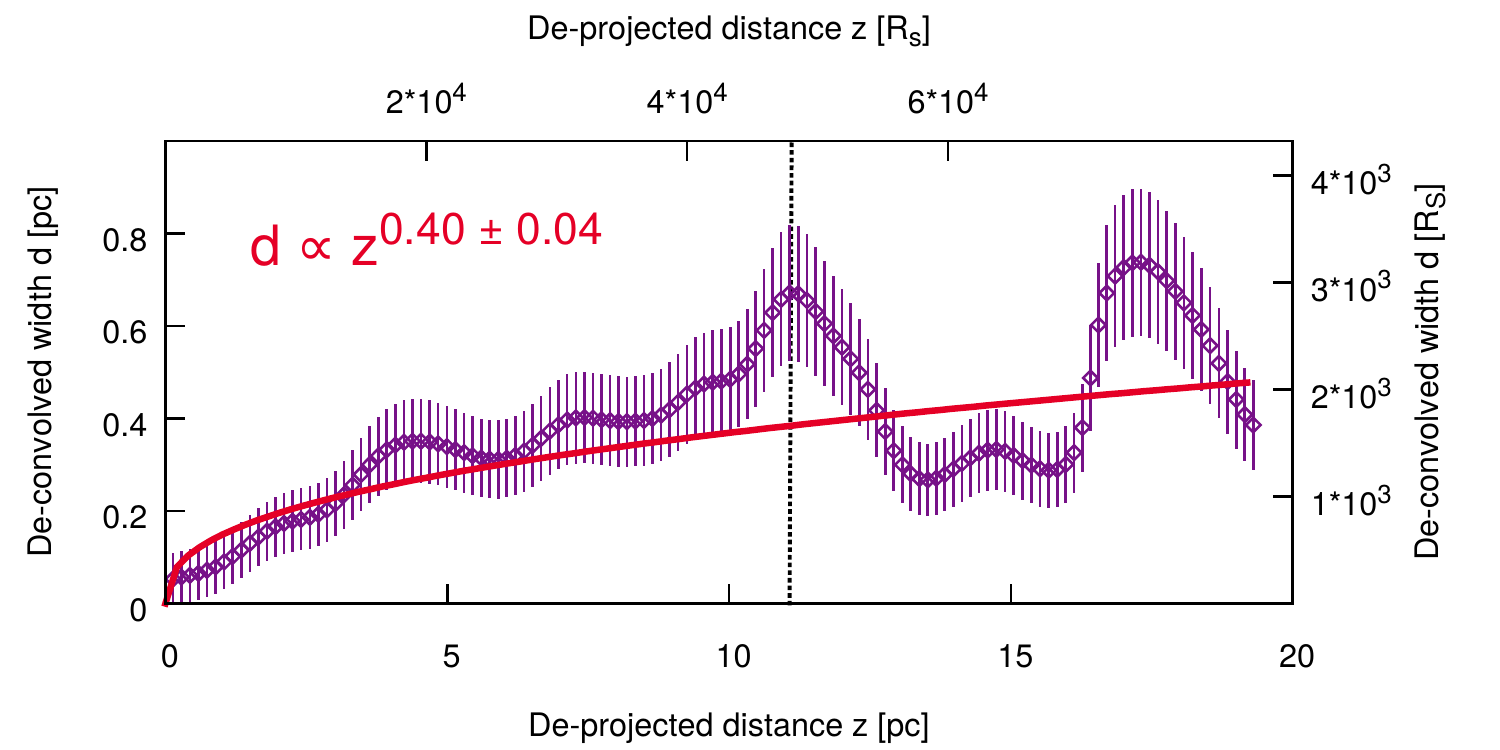}
    \caption{Collimation profile of the jet in 3C\,264. The flow has a close-to-parabolic shape on parsec scales. A recollimation is observed to start at a distance of ${\sim}11$\,$\rm pc$ (dashed vertical line). This location may coincide with the end of the acceleration zone, as seen in Fig. \ref{speed}. A fit performed by excluding the recollimation region, i.e. in the interval $0<z<11$\,$\rm pc$, yields $d\propto z^{0.66\pm 0.02}$.}
    \label{col}
\end{figure*}

\subsection{Jet orientation, Lorentz and Doppler factors}
The detection of moderate-to-high superluminal speeds in the jet of 3C\,264 is in agreement with previous optical kinematic studies of the source \citep{2015Natur.521..495M, 2017Galax...5....8M}, and constrains the jet viewing angle $\theta$ to assume relatively small values. Since we measure a maximum apparent speed of ${\sim}11.5\,\rm{c}$, the equation that relates the apparent to the intrinsic speed,
\begin{equation}
\beta_{\rm app}=\frac{\beta\sin\theta}{1-\beta\cos\theta},
\end{equation}
limits the viewing angle to a maximum value $\theta_{\rm max}\sim10^{\circ}$. On the other hand, given the absence of prominent radio flux density variability (see Table 1), and considering the quite extended large scale morphology of 3C\,264, the viewing angle is unlikely to assume very small values. Therefore, in the following calculations we will assume the upper limit $\theta=10^{\circ}$.  For this orientation, an apparent speed $\beta_{\rm app}=10\,\rm{c}$, which is the average speed in the upper envelope in Fig. \ref{speed}, implies an intrinsic speed $\beta=0.9978$ and a Lorentz factor $\Gamma=1/\sqrt{(1-\beta^2)}\sim15$. For the lower speed envelope with $\beta_{\rm app}=5\,\rm{c}$, instead, we obtain $\beta=0.9808$ and $\Gamma\sim5$. The assumption of the above parameters finally enables us to calculate the Doppler factor $\delta$ of the two envelopes, which is a function of both the intrinsic speed and the orientation:
\begin{equation}
\delta=\frac{1}{\Gamma(1-\beta\cos\theta)},
\end{equation}

For the high and the low speed values we obtain respectively $\delta\sim4$ and $\delta\sim6$. We note that if the slow envelope is actually reflecting the speed of the limbs and the fast one is associated to regions close to the jet axis, the difference between these Doppler factors naturally explains the observed limb-brightening. Moreover, the jet orientation is such that a large difference in $\Gamma$ ($\Gamma\sim15$ versus $\Gamma\sim5$) translates into a moderate difference in $\delta$, which enables the fast filaments to be also visible. In other words, the central filaments are not de-boosted ($\delta<1$), but are less boosted than the limbs. \\

 \subsection{Collimation profile}
 As a final step in the study of the innermost radio jet in 3C\,264, we have examined its collimation properties. The analysis of the jet shape complements the kinematic study presented in Sect. 4.1, as it allows us to further test our finding that bulk acceleration is taking place on the examined VLBI scales. In fact, according to theoretical models, acceleration through MHD processes cannot take place at large distances from the black hole if the jet is freely expanding, i.e., it has a conical shape; therefore a jet which is still accelerating on parsec scales should undergo active collimation in the same regions \citep[see, e.g.,][]{2006MNRAS.367..375B,2007MNRAS.380...51K,2009ApJ...698.1570L}. 
 
 With the aim of determining the jet width at each pixel and investigating its evolution with distance from the core, we have created a new stacked image after restoring all maps with a circular beam of $0.6\times0.6$\,$\rm mas$, corresponding to the average equivalent beam $b_{\rm eq}=\sqrt{b_{\rm min}\times b_{\rm max}}$. The choice of a slightly larger restoring beam in the transverse direction makes the analysis less sensitive to the the jet transverse stratification. The stacked image was sliced pixel by pixel (1px=0.06 $\rm mas$) in the direction perpendicular to the the jet axis using the \textsc{AIPS} task \textsc{SLICE}. Through the task \textsc{SLFIT}, the transverse intensity distribution in each slice was then modeled by fitting a Gaussian profile, in order to extract its width (FWHM) and its peak flux density. 
 
 Figure \ref{col} shows the dependence of the de-convolved jet width $d$ on the de-projected distance from the core $z$. The distance is now expressed in linear scale in units of parsecs and of Schwarzschild radii $R_{\bf S}$, and was de-projected assuming a viewing angle $\theta=10^{\circ}$, as determined in Sect. 4.2. In the plot we only report measurements obtained for transverse profiles with a signal-to-noise ratio higher than 10. The error bars on the width are conservatively assumed equal to one fifth of the convolved FWHM. The jet appears to expand quite smoothly up to a distance of ${\sim}11$\,$\rm pc$ (${\sim}4$ projected $\rm mas$), where a recollimation occurs. By fitting a power law of the form $d=a\cdot z^b$ to the whole profile, we obtain $d\propto z^{0.40\pm 0.04}$. Therefore, as expected for an accelerating jet, the flow is collimating and has a close-to-parabolic shape. This conclusion is still valid if, in the fit, we consider only the inner regions of the jet, excluding the recollimation at ${\sim}11$\,$\rm pc$. In this case, we obtain $d\propto z^{0.66\pm 0.02}$. We note that the distance at which the recollimation occurs coincides with the distance, in Fig. \ref{speed}, where the speeds profile starts flattening, and where the jet limbs intersect (Fig. \ref{vec}). Ultimately, our results support a scenario in which the acceleration and collimation of the jet in 3C\,264 take place in the inner ${\sim}4$ projected milliarcseconds, corresponding to a de-projected distance of ${\sim}11$ parsecs, or ${\sim}4.8\times10^4$\,$R_{\rm S}$ (assuming $M_{\rm BH}\sim4.7\times10^8$\,$M_{\odot}$).

\section{SED modeling}

\begin{table*}
\centering
\caption{Model parameters (see Sec. 5 in the text) and jet powers of the soft low state (SLS) and hard high state (HHS) of 3C~264: L$_{rad}$ is the total (synchrotron and SSC) radiative power, L$_B$ the Poynting flux, L$_e$ and L$_p$ the kinetic powers in emitting electrons and in protons (assuming one cold proton per electron), respectively, and L$_{kin}$=L$_e$+L$_p$.}
\label{SEDpar}
\begin{tabular}{l| c c| c c c }
\hline
\hline
\multicolumn{1}{l|}{}&
\multicolumn{2}{c|}{SLS}&
\multicolumn{3}{c}{HHS}\\
\hline
\multicolumn{1}{l|}{Model}&
\multicolumn{1}{c}{Core}&
\multicolumn{1}{c|}{Core}&
\multicolumn{1}{c}{Core}&
\multicolumn{1}{c}{Layer}&
\multicolumn{1}{c}{Spine}\\
\multicolumn{1}{l|}{Parameters}&
\multicolumn{1}{c}{Model 1}&
\multicolumn{1}{c|}{Model 2\&3}&
\multicolumn{1}{c}{Model 1}&
\multicolumn{1}{c}{Model 2}&
\multicolumn{1}{c}{Model 3}\\
  $\Gamma_{bulk}$ & 2.0 & 2.0 & 2.0 & 5.0 & 8.0\\
  $\theta$ & 10.0 & 10.0 & 10.0 & 10.0 & 5.0\\
  $B$ (G) & 0.055 & 0.12 & 0.062 & 0.0075 & 0.0035\\
  $B_{eq}$ (G) &0.09  &0.04      &0.15  & 0.03        &0.023\\
  $R$ (cm) & 6.5$\times 10^{16}$ & 3$\times 10^{17}$ & 2.3$\times 10^{16}$ & 1.15$\times10^{17}$ & 7$\times 10^{16}$\\
  $\gamma_{min}$ & 2$\times 10^3$ & 2$\times 10^2$ & 2$\times 10^3$ & 3$\times 10^3$ & 3$\times 10^3$\\
  $\gamma_{max}$ & 1$\times 10^6$ & 4$\times 10^5$ & 3$\times 10^6$ & 2$\times 10^6$ & 2$\times 10^6$\\
  $\gamma_{break}$ & 2$\times 10^4$ & 4$\times 10^3$ & 8.5$\times 10^4$ & 3.5$\times 10^3$ & 5$\times 10^3$\\
 $p_1$ & 2.2 & 2.2 & 2.1 & 2.2 & 2.1\\
  $p_2$ & 3.1 & 3.1 & 3.0 & 2.7 & 2.66\\
%  $k$ & 9$\times 10^2$ & 10.0 & 3.5$\times 10^3$ & 2$\times 10^3$ & 1$\times 10^3$\\
  U$_B$/U$_e$ & 0.13 & 37.0 & 0.021 & 0.002 & 0.0003\\
 
  \hline
 Powers   &&&&&\\ 
  L$_{rad}$ (erg/s) & 2.4$\times 10^{41}$ & 3.1$\times 10^{41}$ & 5.1$\times 10^{41}$ & 6.0$\times 10^{41}$ & 1.3$\times 10^{41}$\\
  L$_B$ (erg/s) & 1.7$\times 10^{41}$ & 1.7$\times 10^{43}$ & 2.6$\times 10^{40}$ & 6.8$\times 10^{40}$ & 1.4$\times 10^{40}$\\
  L$_e$ (erg/s) & 5.3$\times 10^{41}$ & 2.3$\times 10^{41}$ & 6.3$\times 10^{41}$ & 1.5$\times 10^{43}$ & 2.1$\times 10^{43}$\\
  L$_p$ (erg/s) & 1.6$\times 10^{41}$ & 6.3$\times 10^{41}$ & 1.4$\times 10^{41}$ & 3.9$\times 10^{42}$ & 5.0$\times 10^{42}$\\
  L$_{kin}$ (erg/s) & 7.0$\times 10^{41}$ & 8.6$\times 10^{41}$ & 7.7$\times 10^{41}$ & 1.9$\times 10^{43}$ & 2.6$\times 10^{43}$\\  
\hline\end{tabular}
\end{table*}

\begin{figure}
\centering 
\includegraphics[trim=1.cm 2.5cm 9cm 7.cm, clip=true, width=0.5\textwidth]{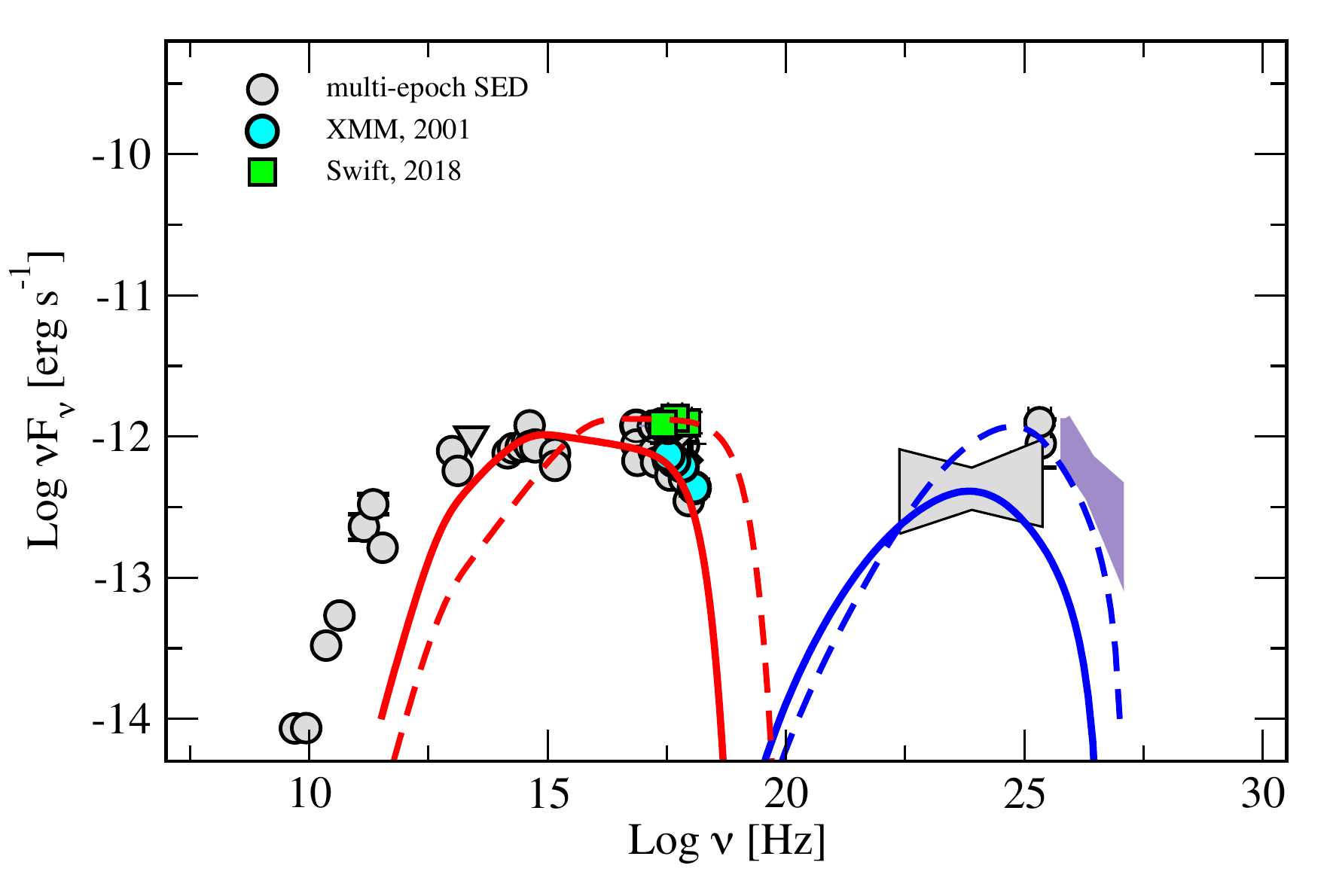}
\includegraphics[trim=1.cm 2.5cm 9cm 7.cm, clip=true, width=0.5\textwidth]{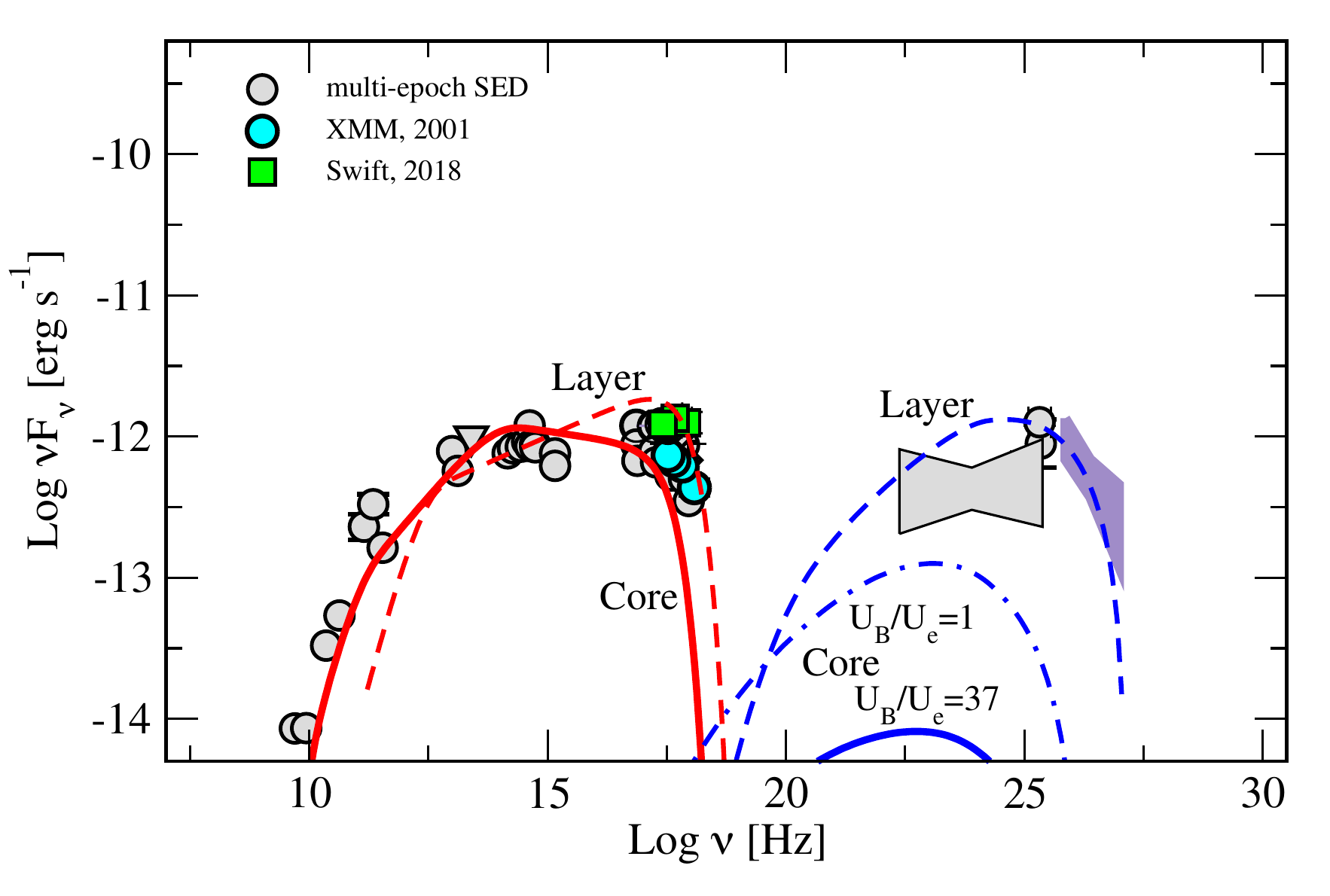}
\caption{Observed SED of 3C\,264 and models. Multi-epoch gray points were retrieved from publicly available databases (NED, ASDC). References for the fluxes are: \cite{2004A&A...415..905L,quillen} for the radio and microwave data, \cite{2009ApJ...701..891L} for infrared data, and \cite{chiaberge, UV} for the optical-UV. The X-ray points are from this paper, while the $\gamma$-ray data are from the 3FGL and 3FHL catalogs \citep{2015ApJ...810...14A,2017ApJS..232...18A}. The TeV spectrum was derived assuming a power law photon index ranging between 2.3 and 2.8 \citep{2018Galax...6..116R} and a flux above 300 GeV as reported by \cite{2018ATel11436....1M}.
Upper panel: the SLS (solid red line for the synchrotron curve and blue line for the SSC component) and HHS (dashed red and blue lines) are both produced by an emitting region in the core (Model 1), still within the acceleration region. A large particle dominance is required to reach the observed \g-ray fluxes. 
Lower panel: the emission of the SLS and HHS are produced in two different regions, in the core and in the layer, respectively (Model 2). This time, the core region is assumed to be Poynting flux-dominated (U$_B$/U$_e$=37). The SSC emission of the core component (solid blue line) is significantly below the observed \g-ray emission, and remains such when equipartition between particles and magnetic field is assumed (dot-dashed line). The HHS originates in a region in the layer at the end of the acceleration region, moving with a bulk motion $\Gamma_{bulk}\sim$5.}
\label{SED}
\end{figure}

In light of the properties inferred from the analysis of the radio and X-ray data, we modeled the nuclear SED of 3C\,264 (Fig. \ref{SED}). 
The radio to \g-ray data were collected from publicly available databases (NED\footnote{https://ned.ipac.caltech.edu/}, ASDC\footnote{https://www.asdc.asi.it/}). In addition to our estimated X-ray fluxes, the average \g-ray flux reported in the 3FGL (light-grey bow tie in Fig. \ref{SED}) and the 3FHL data points are shown. While a detailed analysis of the source behavior in the GeV-TeV regime is beyond the scope of this work, for guidance in the modeling, we included in the SED an estimate of its VHE flux. We used the integrated ($>$300 GeV) flux, ($1.3\pm0.2$)$\times 10^{-12}$\,$ph$ \,$cm^{-2} s^{-1}$ \citep{2018ATel11436....1M}, and assumed a power-law spectral shape with a hard and a steep photon index \citep[$\Gamma_{VHE}=$2.3 and 2.8, see e.g. the review by ][]{2018Galax...6..116R}.
 
The SED displays the characteristic double-hump shape, with the low-energy component extending up to the X-rays and showing no evidence of features related to the accretion process \citep[such as a bump in the UV regime and broad emission lines in the optical band, see also][]{2009A&A...495.1033B, 2010A&A...509A...6B}. Such properties are in agreement with the expectations for a low-power AGN jet which, according to our estimate of the viewing angle, lies at the boundary between FR\,I and BL Lac sources.
The fact that 3C\,264 is not always significantly detected on monthly timescales by {\it Fermi}-LAT and the recent detection at TeV energies suggest that the source is a relatively faint \g-ray emitter with possible, low-amplitude variability \citep[][]{2018Galax...6..116R}.
As discussed in Sect. 3.2, flux variability is clearly observed at X-rays, a property which is also evident at the respective range in the SED. 

In the modeling presented in the following we explore a scenario in which the source switches between two states: a soft/low state (SLS) characterized by the lowest flux and softest X-ray spectrum, and a hard/high state (HHS) when the source is the brightest and has a hard X-ray spectrum. Based on the results obtained in our kinematic study, the modeling relies on two working hypotheses: 
\begin{enumerate}
\item the SLS corresponds to the core's average SED and
\item the core is located at the base of the acceleration region, where the plasma still has a mildly relativistic bulk speed ($\Gamma_b{\sim}2)$. 
\end{enumerate}
In this framework, we will discuss different scenarios for the origin of the HHS.

We assume that the emission is produced in a spherical blob of radius $R$, moving with a bulk Lorentz factor $\Gamma_b$, and seen at a viewing angle $\theta$. The electrons follow a broken power-law energy distribution: $p_1$, $p_2$ are the indices below and above the spectral break, $\gamma_{min}$, $\gamma_{max}$ and $\gamma_{break}$ the minimum, maximum Lorentz factors and the Lorentz factor at the spectral break, respectively. By interacting with the magnetic field ($B$), the electrons radiate via the synchrotron mechanism, which accounts for the low-energy component of the SED. The synchrotron photons are then up-scattered by the same electron population (Synchrotron-Self-Compton, SSC), giving rise to the high-energy hump. 

We first tested a scenario in which the SLS and the HHS emission are both produced in the core (Model 1 - Figure \ref{SED}, upper panel), but the HHS originates in a more compact and more magnetized region.
The model parameters for the two states are shown in Table \ref{SEDpar}. The compact volume assumed for HHS implies that the emission can vary on timescales as short as few days ($t_{\rm var}{\sim}R(1+z)/(c\delta){\approx}2.7$\,$\rm days$), in agreement with those suggested by the Swift monitoring (Sect. 3.2, Fig. 3). In the HHS the blob dissipates almost 60\% of its total power ($L_{jet}=L_{kin}+L_B \sim 8\times 10^{41}$ erg s$^{-1}$) in the form of radiation. 
The model parameters of the SLS are rather similar, suggesting that the latter state could directly derive from the former. The SLS could be the result of strong radiative losses plus adiabatic losses due to the mild volume expansion of the same emitting region (the radius increases by a factor of 2 from the HHS to the SLS).
However, the energetic budget of the two states disfavor this hypothesis, given that the estimated total power in the SLS ($\sim 8.7\times10^{41}$ erg s$^{-1}$) exceeds the one left in the HHS after dissipation. Alternatively, as proposed by some jet models \citep[see e.g.][]{2014ApJ...780...87M}, the two states  could originate in different cells within the radio core region. In the latter case, the estimated powers carried by these emitting regions would be only a fraction of the total jet power, since the jet still has to propagate and form the large scale radio structure. Indeed, a higher jet  power ($L_{jet}=1.8\times10^{44}$ erg s$^{-1}$) was estimated by \cite{2008A&A...486..119B}, based on a semi-empirical relation considering the radio core luminosity \citep{2007ApJ...658L...9H}.

This simple one zone model provides a reasonably good description of the broadband SED in 3C\,264, but it requires the emitting region to be very weakly magnetized. Indeed, as shown in Table 4, the ratio between the energy density of the magnetic field and the leptons, $U_B$ and $U_e$ respectively, is much smaller than unity both in the SLS ($U_B/U_e\sim$0.13) and the HHS ($U_B/U_e\sim$0.02). Similar ratios have been obtained for the misaligned jet in IC~310 \citep{2017A&A...603A..25A}, which is also detected at TeV energies, as well as in a subclass of TeV detected BL Lacs \citep{2010MNRAS.401.1570T} characterized by higher Doppler factors ($\delta\gtrsim 15$). In the case of 3C\,264, this result is in contrast with our initial hypothesis that acceleration of the bulk flow is taking place on the examined VLBI scales. In fact, according to theoretical models and simulations of magnetic jet launching \citep[see e.g.,][and references therein]{2012bhae.book.....M}, in the jet acceleration zone most of the energy should be stored in the magnetic field, and gradually converted into kinetic energy of the bulk flow. In a highly magnetized jet, acceleration of particles radiating up to very high energies may be triggered by magnetic dissipation through reconnection, a process which  requires the field and the emitting particles to be at least in equipartition \citep{2009MNRAS.395L..29G, 2015MNRAS.450..183S}. While a production of high-energy particles in the vicinity of the black hole magnetosphere may still be possible, as suggested by recent magnetospheric models drawn by pulsar-like acceleration mechanisms \citep[see][]{2014Sci...346.1080A,2016ApJ...818...50H,2017A&A...603A..25A}, in the following we test the acceleration scenario by imposing that $U_B/U_e\geq 1$ in the core region (Model 2, Figure \ref{SED}, bottom panel).

With this constrain, the synchrotron emission from the core can still reproduce the radio to X-ray observed SED. However, for $B\gtrsim$0.1 G the SSC curve significantly under predicts the emission at higher energies, even in the most favorable case of equipartition between $U_e$ and $U_B$. Therefore, we consider a case in which the low- and high-energy components originate in two separate emitting regions.
The low-energy synchrotron emission is still produced in an accelerating, Poynting flux-dominated region in the core and accounts for the SLS in X-rays. The variable high-energy emission is instead associated to a blob at the end of the acceleration region ($z{\sim}11$\,$\rm pc$), either in the layer ($\Gamma_b=$5, Model 2) or in the spine ($\Gamma_b=$8-10, Model 3). 
In Model 3, we assume a blob of the spine that is temporarily pointing towards us ($\theta\leq$5\dg). This possibility finds support in the results of kinematics study, showing that some jet features move along curved trajectories, and could account for the fact that the source is only sporadically detected in the GeV to TeV band, as well as in X-rays in the hard state. Variations of the Doppler factor due to a change of $\theta$ have been proved able to consistently reproduce the long-term flux variability of other blazars \citep[e.g. 4C 38.41, CTA 102][and references therein]{2012A&A...545A..48R,2017Natur.552..374R}. 
In Figure \ref{SED}, we show that this scenario can explain even the VERITAS detection in the TeV band and, with some variation of the layer (or spine) parameters the average GeV flux measured by {\it Fermi}.

The total power in the layer (or spine) component is in rough agreement with that of the core (a few $10^{43}$ erg s$^{-1}$), but the Poynting flux, which is energetically dominant in the latter, has been converted to kinetic power in the former. Here, for comparison with previous works, we have included in the jet energy budget a proton component ($L_p$) and assumed one cold proton for radiating electron. However, $L_p$ is not energetically dominant in any of the models, nor strictly necessary, hence a ``light’’ electron-positron jet could be equally possible.

The total power estimated in Model 2 and Model 3 is consistent with the one independently derived by \cite{2008A&A...486..119B}.

\section{Summary}
Motivated by the recent detection of 3C\,264 in the TeV regime \citep{2018ATel11436....1M}, we have conducted a multi-band study of its misaligned jet, combining VLBI radio images with high frequency data collected by different satellites (Chandra, XMM, Swift, Fermi-LAT) and adopting the inferred properties to model the broadband SED. 
\begin{itemize}
\item By employing a multi-scale wavelet decomposition method \citep[WISE code,][]{2015A&A...574A..67M, 2016A&A...587A..52M}, we have investigated the bulk motion within the radio jet, which appears strongly limb-brightened. Superluminal apparent speeds up to ${\sim}11.5$\,$\rm c$ are determined in our analysis. Such high speeds, in agreement with findings from optical kinematics on kilo-parsec scales \citep{2015Natur.521..495M, 2017Galax...5....8M}, limit the viewing angle to relatively small values ($\theta\lesssim10^{\circ}$).  A trend of increasing speed with distance from the core is suggested on a de-projected scale of ${\sim}11$\,$\rm pc$ (or  ${\sim}4.8\times10^4$\,$R_{\rm S}$). The plasma features are observed to move mainly along the jet limbs, but significant transverse motion is also detected when the jet filaments appear to intersect and cross each other. In agreement with theoretical models, which predict active collimation to take place along the acceleration region, the jet has a parabolic shape on the examined VLBI scales ($r\propto z^{0.40\pm 0.04}$). 
\vspace{0.15cm}
\item We have analyzed all the available X-ray data, including those obtained in 2018 through a dedicated Swift monitoring. The X-ray lightcurves indicate the presence of both long and short time variability.  The luminosity increased almost by a factor of ${\sim}3$ from 2000 to 2018, while the spectrum became harder. The Swift data reveal day-scale variability by a factor of ${\sim}2$ between January and April 2018, and a possible flaring event two days after the end of the VERITAS campaign, in March 2018.
\vspace{0.15cm}
\item Based on the radio and X-ray results, we have modeled the broad-band emission assuming that the source switches between a low/soft and a high/hard state. A simple one-zone synchrotron self-Compton model provides a reasonably good description of the SED in both states, but requires the core region to be strongly particle-dominated {\bf ($U_B/U_e{\sim}0.02-0.13$)}. Since we expect the plasma to be still magnetically dominated in the acceleration and collimation region, we have then tested a second scenario by imposing that $U_B/U_e\geq 1$ in the core. The SED, as well as the expected total jet power, can be well reproduced with this assumption, provided that the high-energy component is associated to a second emission region. This could be located, for instance, at the end of the acceleration region, either in the spine or in the sheath.
\end{itemize}

A location of the high-energy emitting region in the jet acceleration zone has been considered in previous studies \citep[e.g.,][]{2012MNRAS.424L..26G} but, for the assumed jet geometry and kinematics, the GeV-TeV emission was found to be significantly suppressed by \g\g-pair production. The results of our kinematic study, however, indicate that the jet acceleration region in 3C\,264 is relatively extended, which allows us to release the constraints on the size and location of the emitting region. There is, on the other hand, increasing evidence, based on VLBI observations, for the existence of such extended acceleration regions. For instance, about half of the AGN jets in the MOJAVE sample presents features moving on accelerating trajectories \citep{2016AJ....152...12L}, on linear scales reaching up to hundreds of parsecs. In the case of 3C\,264, its proximity and the relatively large black hole mass and viewing angle enable us to probe the innermost jet with high resolution in units of Schwarzschild radii, on scales where the bulk of the MHD processes is still taking place.

 \begin{acknowledgements} 
 The authors would like to thank the anonymous referee for the useful comments. We also thank Carolina Casadio for reading the paper and for her suggestions. This research has made use of data from the MOJAVE database that is maintained by the MOJAVE team \citep{2018ApJS..234...12L}. 
 \end{acknowledgements}
 
\bibliographystyle{aa}
\bibliography{reference.bib}
\end{document}